\documentclass[aps,11pt,notitlepage,nofootinbib,showkeys,prd]{revtex4-1}
\usepackage{amstext,amsmath,amssymb,amsfonts,bbm}
\usepackage[latin1]{inputenc}
\usepackage{epsfig}
\usepackage{hyperref}
\usepackage{amsthm}
\usepackage{subfigure}
\usepackage{color}
\usepackage{multirow}
\newcommand{\bea}{\begin{eqnarray}}	
\newcommand{\eea}{\end{eqnarray}}
\newcommand{\cG}{{\cal G}}

\newcommand{\cF}{{\cal F}}
\newcommand{\cB}{{\cal B}}
\newcommand{\cT}{{\cal T}}

\newcommand{\cL}{{\cal L}}
\newcommand{\cN}{{\cal N}}

\newcommand{\cM}{{\cal M}}

\newcommand{\cJ}{{\cal J}}

\DeclareMathOperator{\tr}{Tr}

\newtheorem{lemma}{Lemma}

\newtheorem{definition}{Definition}
\newtheorem{theorem}{Theorem}

\begin{document}

\title{\Large \bf A generalization of the Virasoro algebra to arbitrary dimensions}

\author{{\bf Razvan Gurau}}\email{rgurau@perimeterinstitute.ca}
\affiliation{Perimeter Institute for Theoretical Physics, 31 Caroline St. N, ON N2L 2Y5, Waterloo, Canada}

\begin{abstract}
  Colored tensor models generalize matrix models in higher dimensions. They admit a 1/N expansion 
dominated by spherical topologies and exhibit a critical behavior strongly reminiscent of matrix models. 
In this paper we generalize the colored tensor models to colored models with generic interaction, derive  
the Schwinger Dyson equations in the large $N$ limit and analyze the associated algebra of constraints 
satisfied at leading order by the partition function. We show that the constraints form a Lie 
algebra (indexed by trees) yielding a generalization of the Virasoro algebra in arbitrary dimensions. 
\end{abstract}

\medskip

\noindent  Pacs numbers: 02.10.Ox, 04.60.Gw, 05.40-a

\maketitle

\section{Introduction}

Random matrix models \cite{mm,david-revueDT,Di Francesco:1993nw} generalize in higher dimensions to 
random tensor models \cite{mmgravity,ambj3dqg,sasa1,ambjorn-book, ambjorn-houches94} and group field 
theories \cite{Boul,Ooguri:1992eb,laurentgft,quantugeom2} (see 
\cite{dev1,dev2,dev3,dev4,dev5,dev6} for some further developments). 
The Feynman graphs of GFT in $D$ dimensions are built from vertices dual to 
$D$ simplices and propagators encoding the gluing of simplices along their boundary. 
Parallel to ribbon graphs of matrix models (dual to discretized surfaces), GFT 
graphs are dual to discretized $D$ dimensional topological spaces.
Tensor models are notoriously hard to control analytically and one usually resorts to
numerical simulations \cite{Ambjorn:1991wq,Ambjorn:2000dja,Ambjorn:2005qt}.
Progress has recently been made in the analytic control of tensor models with the advent of the
$1/N$ expansion \cite{Gur3,GurRiv,Gur4} of {\em colored} \cite{color,lost,PolyColor} tensor models. 
This expansion synthesizes several evaluations of graph amplitudes 
\cite{FreiGurOriti,sefu1,sefu2,sefu3,BS1,BS2,Geloun:2011cy,BS3,Carrozza:2011jn}
and provides a straightforward generalization of the familiar genus expansion of matrix models
\cite{'tHooft:1973jz,Brezin:1977sv} in arbitrary dimension.
The coloring of the fields allows one to address previously inaccessible questions like the
implementation of the diffeomorphism symmetry
\cite{Baratin:2011tg,Girelli:2010ct,Carrozza:2011jn} or the identification of embedded matrix models \cite{Ryan:2011qm}
in tensor models. The symmetries of generic tensor models have recently been studied using n-ary algebras
\cite{Sasakura:2011ma,Sasakura:2011nj}. 

The critical behavior of matrix models is most conveniently addressed using the loop equations 
\cite{Ambjorn:1990ji,Fukuma:1990jw,Makeenko:1991ry} in conjunction with the $1/N$ expansion.
The loop equations translate in a set of constraints (obeying the Virasoro algebra) 
satisfied by the partition function and provide the link between matrix models and
continuum conformal field theories. Generic matrix models exhibit multi critical 
points \cite{Kazakov:1989bc} which are at the core of their applications 
to string theory \cite{Gross:1990ay,Gross:1990ub}, two dimensional gravity \cite{Di Francesco:1993nw}, 
critical phenomena \cite{mm,Kazakov:1985ea,Boulatov:1986jd}, black hole physics \cite{Kazakov:2000pm}, etc. 

Recently the investigation of the critical behavior of the simplest colored tensor models has been performed
\cite{Bonzom:2011zz} mapping  the dominant family of graphs (generalizing the planar \cite{Brezin:1977sv} 
graphs of matrix models) on certain species of colored trees.

However, up to now, the colored tensor models considered posses only one interaction term (one coupling constant). 
It is well known that multi critical points for matrix models appear only when one adds multiple interaction terms.
A first question we will solve in this paper is to write a colored tensor model with generic interactions. 
In order to access the critical behavior of such a model one must derive a generalization of the loop 
equation in higher dimensions. We will derive the closed set of Schwinger-Dyson Equations (SDE) 
obeyed by a generic colored tensor models in the large $N$ limit which we translate in constraints on 
the partition function. We will prove that the constraints form a Lie algebra yielding a higher 
dimensional generalization of the Virasoro algebra.

This paper is organized as follows. In section \ref{sec:matrixmodels} we recall the derivation 
of the loop equations and the link with the Virasoro algebra in matrix models. 
In section \ref{sec:algebra} we introduce an algebra indexed by colored trees. 
In section \ref{sec:SDE} we derive the SDEs at leading
order in the $1/N$ expansion of colored tensor models and translate them into 
constraints on the partition function which we identify with the 
generators of our algebra. Section \ref{sec:conclu} draws the conclusions of this work.

\section{Matrix Models and the Virasoro algebra}\label{sec:matrixmodels}

This section is a quick digest of \cite{Ambjorn:1990ji,Makeenko:1991ry} and presents the classical 
derivation of the loop equations and their link with the Virasoro algebra in matrix models.
In the spirit of our subsequent treatment of colored tensor models, we will start from a colored matrix model
\cite{difrancesco-rect,difrancesco-coloringRT,difrancesco-countingRT} of three independent non hermitian matrices $M_1, M_2$ and $M_3$, defined 
by the partition function 
\bea
&& Z = \int [dM_1] [dM_2] [dM_3 ] \quad e^{-N \tr[ V(M_1,M_2,M_3)]},\crcr
&&V(M_1,M_2,M_3) = M_1\,M_1^\dagger + M_2\,M_2^\dagger + M_3\,M_3^\dagger 
- \lambda M_1M_2M_3 - \bar \lambda M_3^\dagger M_2^\dagger M_1^\dagger \; ,
\eea
with $[dM] = \prod_{a,b} dM_{ab} d\bar M_{ab}$. 
As the integral is Gaussian, one can explicitly integrate over two colors to obtain
the partition function as an integral over one matrix
\bea
&& Z = \int [dM_3] \quad e^{-N \tr[V(M_3M_3^{\dagger})]} \crcr
&& V(M_3M_3^\dagger) = M_3M_3^{\dagger} + \ln (\mathbb{I} - \lambda \bar\lambda \; M_3M_3^{\dagger})
 = M_3M_3^{\dagger} + \sum_j \frac{(\lambda\bar\lambda)^j}{j} (M_3M_3^{\dagger})^j \; .
\eea
To pass from a model with one coupling constant to a generic matrix model, one attributes to 
every operator in the effective action for the last color an independent coupling constant
replacing $\frac{(\lambda\bar\lambda)^j}{j}$ by $t_j$
\bea
 Z = \int [dM] e^{-N \tr[V(MM^{\dagger}) ]} \; , \qquad V (M M^{\dagger}) = \sum_{j=1} t_j \; (MM^{\dagger})^j \; .
\eea
The Schwinger-Dyson equations (SDE) of a generic matrix model write
\bea
 0&=& \int [dM] \; \frac{\delta}{\delta M_{ab}} \Big{(} [(MM^{\dagger})^n M]_{ab} \; e^{-N\tr[V(MM^{\dagger})] }\Big{)} \crcr
&=& \Big{\langle} \sum_{k=0}^n [(MM^{\dagger})^{k}]_{aa} [(M^{\dagger}M)^{n-k}]_{bb} \Big{\rangle} 
-N \Big{\langle} \sum_{j=1} j \; t_j [(MM^{\dagger})^n M]_{ab} [M^{\dagger}(MM^{\dagger})^{j-1}]_{ba} \Big{\rangle} \; ,
\eea
which, summing over $a$ and $b$, becomes
\bea
 \Big{\langle} \sum_{k=0}^n \text{Tr}[(MM^{\dagger})^k]\text{Tr}[(MM^{\dagger})^k]^{n-k} \Big{\rangle}
 -N \sum_{j} j\; t_j \Big{\langle} \text{Tr}[(MM^{\dagger})^{n+j}]\Big{\rangle} =0 \; .
\eea
Every insertion of an operator $\tr[(MM^{\dagger})^j]$ in the correlation function can be re expressed as a derivative 
of $V(MM^{\dagger})$ with respect to $t_j$. Consequently the SDEs become 
\bea\label{eq:virmat}
 L_n Z &=& 0 \;, \text{ for } n\ge 0 \; ,\crcr
 L_{n} &=& N^2 \delta_{0,n} - \frac{2}{N} \frac{\partial }{\partial t_n} + 
\frac{1}{N^2}\sum_{k=1}^{n-1} \frac{\partial^2}{\partial t_{k}\partial t_{n-k}} +
  \sum_{j=1}^{\infty} j\; t_j \frac{\partial }{\partial t_{n+j}} \; ,
\eea
where the derivatives w.r.t. $t_j$, with  $j\le 0$ are understood to be omitted.
A direct computation (involving some relabeling of discrete sums) shows \cite{Fukuma:1990jw} 
that the $L_n$'s respect the commutation relations of (the positive operators of) the Virasoro algebra
\bea
 [L_m,L_n]= (m-n) \; L_{m+n} \; \text{ for } m,n\ge 0 \;.
\eea
Note that as we only deal with $L_m,\; m\ge 0$ we of course do not obtain the central charge term.
The key to recovering the Virasoro algebra is the presence of the last term in 
eq. \eqref{eq:virmat}. The truncated operators 
$L_{m}' = L_{m} - \frac{1}{N^2}\sum_{k=1}^{n-1} \frac{\partial^2}{\partial t_{k}\partial t_{n-k}}$
also respect $[L_m',L_n']= (m-n) \; L_{m+n}'$. If one specializes the operators we
define below for $D=2$ (and takes into account the cyclicity of the trace), 
one obtains the operators $L_m'$.

This classical result is our guide towards deriving SDEs and loop equations for 
colored tensor models in arbitrary dimensions. 

\section{A Lie algebra indexed by colored, rooted, D-ary trees}\label{sec:algebra}

As the higher dimensional generalization of the Virasoro algebra we obtain is rather
non trivial we will first present it in full detail and only later identify it with 
the algebra of constraints satisfied by the partition function. 
The operators in our algebra are indexed by colored rooted $D$-ary trees.
Trees and $D$-ary trees are well studied in the mathematical literature \cite{manes-kary-trees}.
The colored rooted $D$-ary trees index the leading order in the $1/N$ expansion of colored tensor models  
\cite{Bonzom:2011zz}.

\subsection{Colored rooted D-ary trees: Definitions}

 A {\bf colored rooted $D$-ary tree} $\cT$ with $|\cT|$ vertices 
is a tree with the following properties
\begin{itemize}
 \item It has a root vertex, denoted $(\;)$, of coordination $D$.
 \item It has $|\cT|-1$ vertices of coordination $D+1$ (i.e. each of them has $D$ descendants). 
 \item It has $(D-1)|\cT|+1 $ leaves of coordination $1$ (i.e. with no descendants).
 \item All lines have a color index, $0,1,\dots ,D-1$, such that the $D$ direct descendants (leaves or vertices)
  of a vertex (or of the root) are connected by lines with different colors.
\end{itemize}

\begin{figure}[htb]
\begin{center}
 \includegraphics[width=4cm]{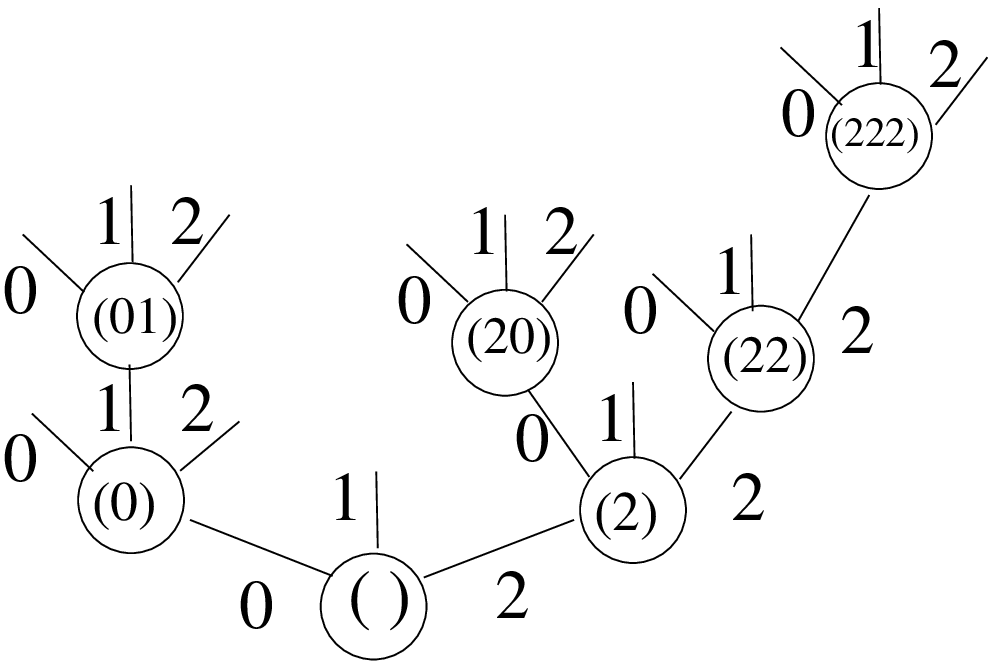}
\caption{A colored rooted $D$-ary tree.}
\label{fig:Dary}
\end{center}
\end{figure}

We will ignore in the following the leaves of the tree, as they can automatically be added once the vertices
and lines of the colored tree are known. A crucial fact in the sequel is that a colored rooted $D$-ary tree
admits a canonical labeling of its vertices. Namely, every vertex can be labeled by the list of colors 
$V=(i_1, \dots, i_n)$ of the lines in the unique path connecting 
$V$ to the root $(\;)$. The first color, $i_1$, is the color of the line in the path ending on the root
(and $i_n$ is the color of the line ending on $V$).
For instance the vertex $(0)$ is the descendant connected to the root $(\; )$ by the line 
of color $0$, and the vertex $(01)$ is the descendant of the vertex $(0)$ connected to it by a line
of color $1$ (see figure \ref{fig:Dary} for an example of a canonically labeled $3$-ary tree with 
$|\cT|=7$ vertices).

In the sequel ``tree'' will always mean a colored rooted $D$-ary tree. A tree is completely identified
by its canonically labeled vertices, hence it is a set $\cT=\{ (\;), \dots\}$.
We denote $\underbrace{i,\dots,i}_{n}$ a list of $n$ identical labels $i$. We will use 
the shorthand notation $V$ for the list of labels identifying a vertex.
Let a tree $\cT$, and one of its vertices 
\bea
V = l, \dots, k , \underbrace{i, \dots, i}_{n} \; ,
\eea
with $k \neq i$. The colors $l,\dots, k$ might be absent.
The {\bf successor of color $j$} of $V$, denoted $s^j_{\cT}[V]$, is the vertex   
\bea
 s^j_{\cT}[ ( l, \dots, k , \underbrace{i, \dots i}_{n }) ]
                = \begin{cases}
                ( l, \dots, k , \underbrace{i, \dots i}_{n}, j) 
                  \text{ if it exists } \\
                \text{ if not }   \begin{cases}
                                   ( l, \dots, k , \underbrace{i, \dots i}_{n }) 
                                               \text{ if } j \neq i \\
                                   ( l, \dots, k ) \text{ if } j=i  
                                  \end{cases}
                 \end{cases} \; .
\eea
The colored successor functions are cyclical, namely if a vertex does not have a descendant in the tree
of color $j$, then its ``successor of color $j$'' is the first vertex one encounters, when going form $V$ to the root,
whose label does not end by $j$. For the example of figure \ref{fig:Dary} we have.
\bea
&& s^0[(\; )]=(0) \quad s^{1}[ (\;) ]=  (\;) \quad s^{2} [(\;)] = (2) \crcr
&& s^0[(0)] = (\;) \quad s^1[(0)]= (01) \quad s^2[(0)] = (0) \crcr 
&& s^0[(01)] = (01) \quad s^1[(01)]= (0) \quad s^2[(01)] = (01) \crcr
&& s^0[(2)] = (20) \quad s^1[(2)]= (2) \quad s^2[(2)] = (22) \crcr
&& s^0[(20)] = (2) \quad s^1[(20)]= (20) \quad s^2[(20)] = (20) \crcr
&& s^0[(22)] = (22) \quad s^1[(22)]= (22) \quad s^2[(22)] = (222) \crcr
&&s^0[(222)] = (222) \quad s^1[(222)]= (222) \quad s^2[(222)] = (\;) \; .
\eea

We call $V$ the {\bf maximal vertex of color $i$} in a tree $\cT$ if
\bea
V = \underbrace{i,\dots i}_{n} \qquad \text{ such that } 
\qquad s^i_{\cT}[ ( \underbrace{i,\dots i}_{n} ) ] = (\;) \; .
\eea
In figure \ref{fig:Dary} the maximal vertex of color $2$ is $(222)$, the maximal 
vertex of color $0$ is $(0)$ and the maximal vertex of color $1$ is the root $(\;)$ itself.

We define the {\bf branch of color $i$ of $\cT$}, denoted $\cT^i$, the tree
\bea
 \cT^i = \Big{\{} (X) \; \vert \; (i,X) \in \cT \Big{\}} \; .
\eea
The branch $\cT^i$ can be empty. The root of the branch $\cT^i$, $(\;)\in \cT^i$ 
corresponds to the vertex $(i) \in \cT$.
The rest of the vertices of $\cT$ (that is the root $(\;)$ and all vertices of the form 
$(k,U)\in \cT,\; k\neq i$) also form  canonically labeled tree $\tilde \cT^i$,
the {\bf complement in $\cT$} of the branch $\cT^i$. 
In figure \ref{fig:Dary}, the branch of color $2$ is the tree
$\cT^2=\{ (\;), (2), (22), (0)\}$, as all the vertices $(2)$, $(22)$, $(222)$
and $(20)$ belong to $\cT$. Its complement is $\tilde \cT^2=\{ (\;), (0), (01)\}$.

\begin{figure}[htb]
\begin{center}
 \includegraphics[width=8cm]{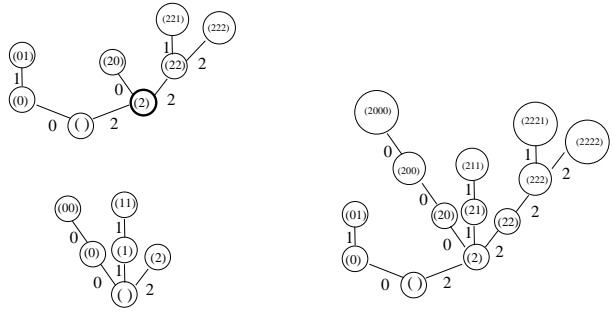}
\caption{Gluing of two trees at a vertex $\cT \star_{(2)} \cT_1$.}
\label{fig:glue}
\end{center}
\end{figure}

Two colored rooted $D$-ary trees $\cT$ and $\cT_1$ can be {\bf joined} (or glued) at a vertex $V\in \cT$. 
For all colors $i$, denote the maximal vertices of color $i$ of $\cT_1$
\bea 
 (\underbrace{i,\dots,i}_{n_i} ) \; ,
 \qquad s^i_{\cT_1}[ ( \underbrace{ i,\dots,i }_{n_i } ) ] = (\;) \; .
\eea 
The glued tree $\cT \star_{V} \cT_1$, is the tree canonically labeled
\bea
 \cT \star_{V}\cT_1 = \begin{cases}
                       (X)     &\text{ for all } (X) \in \cT \;, \;\; (X) \neq (V, \dots) \\ 
                       (V,Y)   &\text{ for all } (Y) \in \cT_1 \\
                       (V, \underbrace{i,\dots,i}_{n_i + 1 } , Z )
                         &\text{ for all } (V,i,Z) \in \cT
                      \end{cases} \; .
\eea
This operation can be seen as cutting all the branches starting at $V$ in $\cT$, gluing the 
tree $\cT_1$ at $V$, and then gluing back the branches at the maximal vertices of the 
appropriate color in $\cT_1$. The vertices of $\cT\setminus (V)$ and $\cT_1\setminus (\;)$ 
map one to one onto the vertices of $ (\cT \star_V \cT_1) \setminus (V)$, and both
$(V)\in \cT, (\;)\in \cT_1$ map to $(V) \in  \cT \star_V \cT_1 $, thus
$ | \cT \star_V \cT_1 | = |\cT| + |\cT_1| -1$.
An example is given in figure \ref{fig:glue}, where the leaves
are not drawn.

\begin{figure}[htb]
\begin{center}
 \includegraphics[width=8cm]{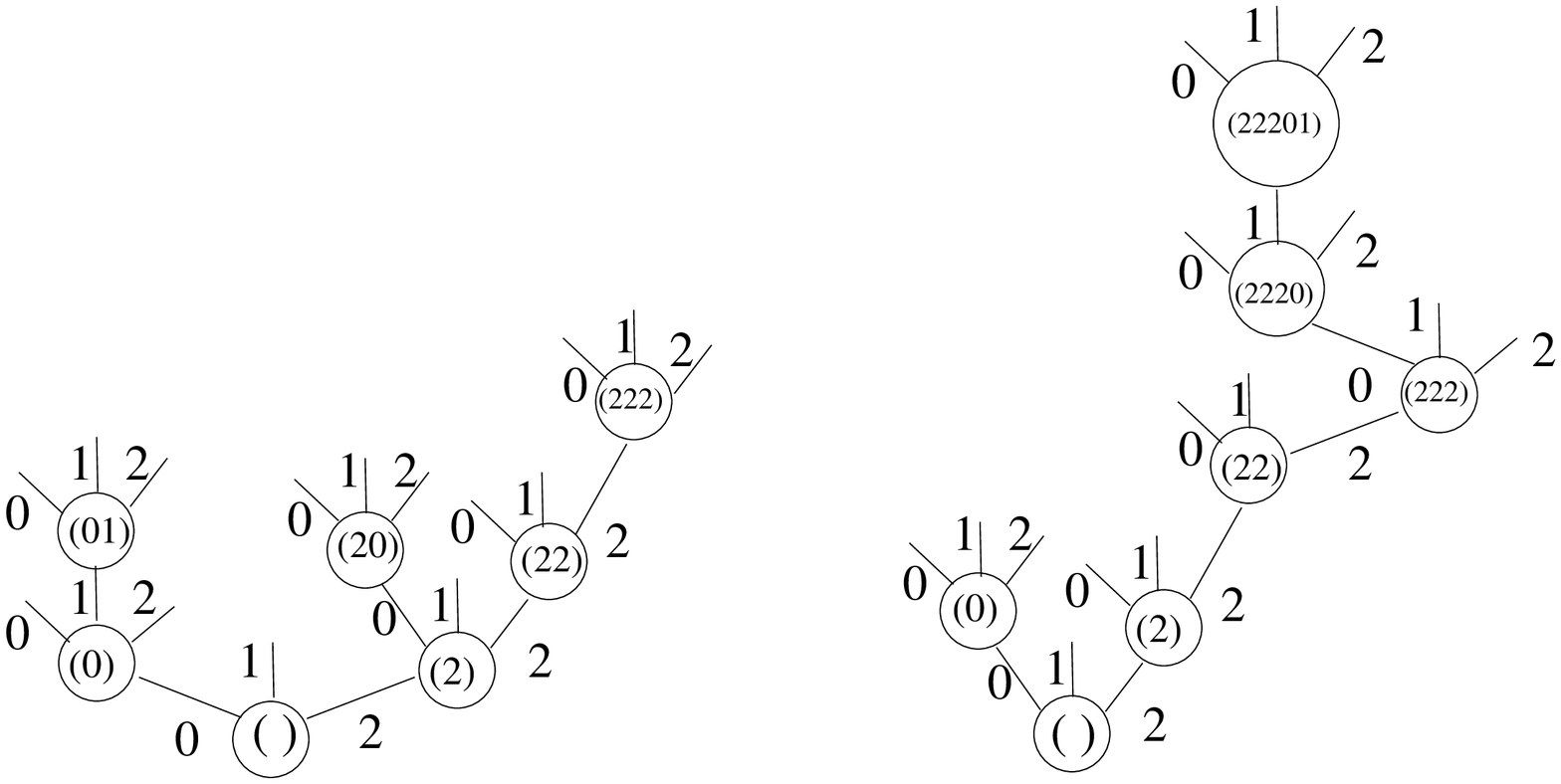}
\caption{Two equivalent trees $\cT \sim \cT'$, with 
$\cT' = \Bigl( \cT^2 \star_{(22)} \{(\;),(2)\} \Bigr) \star_{(222)} \tilde \cT^2$. }
\label{fig:equiDary}
\end{center}
\end{figure}

For any tree $\cT$, with maximal vertex of color $i$, $(\underbrace{i,\dots,i}_{n_i})$,
the maximal vertex of color $i$ in the branch $\cT^i$ will have one less label $(\underbrace{i,\dots,i}_{n_i-1})$.
One can glue the tree $\{ (\;), (i)\}$
on $(\underbrace{i,\dots,i}_{n_i-1} ) \in \cT^i$. All the vertices of $\cT^i$ are unchanged by this gluing,
its only effect being to introduce a new vertex, $(\underbrace{i,\dots,i}_{n_i})$ 
which becomes the new maximal vertex of color $i$. Subsequently, one can glue 
the complement of the branch $i$, $\tilde \cT^i$ on this new maximal vertex 
\bea
 \cT' = \Bigl( \cT^i \star_{ (\underbrace{i,\dots i}_{n_i-1} ) } \{(\;),(i)\} \Bigr)
\star_{ (\underbrace{i,\dots i}_{n_i} ) } \tilde \cT^i \; .
\eea 
The two trees $\cT$ and $\cT'$ have the same number of vertices
$|\cT'| = |\tilde \cT^i|+| \cT^i |=|\cT|$, and the vertices of the initial tree $\cT$ 
map one to one on the vertices of the final $\cT'$. 
As none of the vertices of $\tilde \cT^i$ starts by a label $i$, it follows that 
the maximal vertex of color $i$ in $\cT'$ is
$( \underbrace{i,\dots,i}_{n_i} ) $. Thus
\bea 
 (i,V) \in \cT &\leftrightarrow& (V) \in \cT' \; , \;\; (V)\neq ( \underbrace{i,\dots i}_{n_i} , U ) \;, \crcr
 (W)\in \cT\;, \;\; (W)\neq (i,V) \; &\leftrightarrow& ( \underbrace{i,\dots i}_{n_i} , W ) \in \cT' \; .
\eea
Most importantly, it is straightforward to check that the mapping is consistent with the successor functions
\bea
 V,W \in \cT \leftrightarrow V',W' \in \cT' \text{ such that } s^i_{\cT}[V]=W \Leftrightarrow s^i_{\cT'}[V']=W' \; .
\eea 
We will say that the two trees $\cT$ and $\cT'$ are {\bf equivalent}, $\cT \sim \cT'$, and extended by
transitivity $\sim$ to an equivalence relation between rooted trees.
An example is presented in figure \ref{fig:equiDary}.

The equivalence class of a tree $[\cT]$ has $|\cT|$ members 
all obtained by choosing a vertex $V=i,j,k \dots, l$ in $\cT$ performing
the elementary operation $\sim$ on the colors $i$ followed by $j$ followed by $k$ and so on up to $l$.

\subsection{Colored rooted D-ary trees: Properties}

In this section we prove a number of lemmas concerning the gluing of trees, $\star$. 
All this properties can be readily understood in terms of the graphical representation 
of the trees. In the sequel we will deal with three trees $\cT$, $\cT_1$ and $\cT_2$.
We denote $( \underbrace{i,\dots i}_{n_i} )$ the maximal vertex of color $i$ in $\cT_1$
and $( \underbrace{i,\dots i}_{q_i} )$ the maximal vertex of color $i$ in $\cT_2$.

\begin{lemma} \label{lem:lema1}
 If $ (V) =(k,U) \in \cT $ then  $ (\cT\star_V \cT_1)^k =\cT^k \star_U \cT_1$, 
  and, for $i\neq k$, $ (\cT\star_{V} \cT_1)^i  = \cT^i$.
\end{lemma}
{\bf Proof:} The composite tree $ \cT\star_V \cT_1 $ is
\bea
 \cT\star_V \cT_1 =\begin{cases}
                      (i,X)     &\text{ for all } (i,X) \in \cT \;, \;\; (i,X) \neq (k,U, \dots) \\ 
                      (k,U,Y)   &\text{ for all } (Y) \in \cT_1 \\
                      (k,U, \underbrace{i,\dots,i}_{n_i +1 } , Z )
                         &\text{ for all } (k,U ,i,Z) \in \cT
                   \end{cases} \; .
\eea
It follows that
\bea 
i\neq k: \qquad  (\cT\star_V \cT_1 )^i = \Big{\{} (X) & \text { for all } (i,X)\in \cT\ \Big{\}} =\cT^i \; ,
\eea
and the branch of color $k$ of $\cT\star_V \cT_1$ is
\bea
 (\cT\star_V \cT_1)^k =\begin{cases}
                      (X)     &\text{ for all } (k,X) \in \cT \;, \;\; (k,X) \neq (k,U, \dots) \\ 
                      (U,Y)   &\text{ for all } (Y) \in \cT_1 \\
                      (U, \underbrace{i,\dots,i}_{n_i + 1 } , Z )
                         &\text{ for all } (k,U ,i,Z) \in \cT
                   \end{cases} \; ,
\eea
As $\cT^k = \{ (X) \; | \; (k,X) \in \cT \}$ it follows that $(U) \in \cT^k$ and
\bea
 \cT^k \star_U \cT^1 = \begin{cases}
                        (X) & \text{ for all } (X) \in \cT^k \; , \; \; (X) \neq (U ,\dots )\Leftrightarrow\\
                         & \qquad \qquad \qquad \qquad  \Leftrightarrow
                         (k,X)\in \cT  \; , \; \; (k,X) \neq (k,U ,\dots ) \\
                        (U,Y)  &\text{ for all } (Y) \in \cT_1 \\
                        (U, \underbrace{i,\dots,i}_{n_i +1 } , Z ) &\text{ for all } 
                        (U ,i,Z) \in \cT^k \Leftrightarrow
                        (k,U ,i,Z) \in \cT
                       \end{cases} \; .
\eea

\qed

\begin{lemma}\label{lem:lema2}
  For any three trees, $(\cT \star_V \cT_1) \star_{V} \cT_2= \cT\star_{V} (\cT_1 \star_{(\;)} \cT_2) $.
\end{lemma}
{\bf Proof:} Note that the label of any vertex in $ \cT_1$ starts by the empty label $(\;)$, hence 
the joining at the root writes
\bea
 \cT_1 \star_{(\;)} \cT_2 = \begin{cases}
                             (Y_2) & \text{ for all } (Y_2) \in \cT_2 \\ 
                             (\underbrace{i,\dots, i}_{q_i+1},Y_1) & \text{ for all } (i,Y_1)\in \cT_1
                            \end{cases} \; .
\eea 
The maximal vertices of color $i$ in $ \cT_1 \star_{(\;)} \cT_2 $ are
$(\underbrace{i,\dots, i}_{q_i + n_i}) $. Both operations lead to a tree with vertices
\bea
 \begin{cases}
  (X) &\text{ for all } (X)\in \cT \;, \;\; (X) \neq (V, \dots) \\ 
  (V ,Y_2) &\text{ for all } (Y_2) \in \cT_2 \\
  (V, \underbrace{i,\dots i}_{q_i+1}, Y_1) & \text{ for all } (i,Y_1) \in \cT_1 \\
  (V, \underbrace{i,\dots i}_{q_i+n_i+1}, Z) & \text{ for all }
   (V,i,Z) \in \cT
 \end{cases} \; .
\eea

\qed

\begin{lemma}\label{lem:lema3}
 For any two distinct vertices $V \neq W \in \cT$, 
we denote $ W' $ the image of $W$ in the tree $ \cT \star_V \cT_1 $
and $V' $ the image of $V$ in $ \cT \star_ W \cT_2  $. Then
$(\cT \star_V \cT_1) \star_{W'} \cT_2= (\cT \star_ W \cT_2) \star_{V'} \cT_1$.
\end{lemma}
{\bf Proof:} We distinguish two cases. Either $V$ and $W$ are not ancestor to each other in $\cT$
(both $ (W) \neq (V, \dots)$ and $ (V) \neq (W, \dots) $).
Hence $ W' =W $ and $ V'=V $. In this case,
both operations lead to 
\bea
 \begin{cases}
  (X) & \text{ for all } (X) \in \cT\; , \;\; (X) \neq (V,\dots) \text{ and } (X)\neq (W,\dots) \\
  (V, Y_1) & \text{ for all } Y_1 \in \cT_1 \\
  (V,\underbrace{i,\dots, i}_{n_i+1}, Z ) & \text{ for all } (V,i,Z) \in \cT \\
  (W, Y_2) & \text{ for all } Y_2 \in \cT_1 \\
  (W,\underbrace{i,\dots, i}_{q_i+1}, Z ) & \text{ for all } (W,i,Z) \in \cT
 \end{cases} \; .
\eea
Or one of them (say $V$) is ancestor to the other (hence $(W)=(V,j,A)$ ). 
In this case $ V' =V  $, but 
$ W' = V, \underbrace{j,\dots j}_{n_j+1}, A$.
Both operations lead to the tree
\bea
 \begin{cases}
  (X) &\text{ for all } X \in \cT \;, \; \; X \neq (V, \dots)\\ 
  (V,Y_1) & \text{ for all } Y_1\in \cT_1 \\
  (V,\underbrace{i,\dots i}_{n_i+1} ,Z) & \text{ for all } (V,i,Z) \in \cT \; 
   \text{ with } (V,i,Z) \neq (V,j,A,Z')=(W,Z') \\
  (V, \underbrace{j,\dots j}_{n_j+1}, A, Y_2 ) & \text{ for all } Y_2\in \cT_2 \\
  (V, \underbrace{j,\dots j}_{n_j+1}, A, \underbrace{i,\dots i}_{q_i+1}, Z ) 
   &\text{ for all } (V,j,A,i,Z)=(W,i,Z) \in \cT \\ 
\end{cases} \; .
\eea

\qed

\begin{lemma}\label{lem:lema4}
  Let $V\in \cT$ and $W \in \cT_1$ then 
$(\cT \star_V \cT_1) \star_{W'} \cT_2 = \cT \star_{V} (\cT_1 \star_{W} \cT_2)$, where
again $W'$ denotes the image of $W$ in the tree $ \cT \star_V \cT_1 $.
\end{lemma}
{\bf Proof:} The image of $W$, $ W' = V,W $.
Both joinings lead to the tree
  \bea
     \begin{cases}
      (X) & \text{ for all } X \in \cT  \; X \neq(V,\dots) \\
      (V,Y_1) &\text{ for all } Y_1 \in \cT_1 \; Y_1 \neq(W, \dots) \\
      (V, W, Y_2) & \text{ for all } Y_2 \in \cT_2 \\
      (V, W,\underbrace{i,\dots i }_{q_i + 1} , Z ) & \text{ for all } (W,i,Z) \in \cT_1 \\
      (V ,\underbrace{i,\dots i }_{n_i +1 } , Z) &\text{ for all } (V,i,Z) \in \cT  
                          \text{ if }  (W) \neq (\underbrace{i,\dots,i}_{w_i}) \\
      (V ,\underbrace{i,\dots i }_{q_i + n_i +1 },
          Z) &\text{ for all } (V,i,Z) \in \cT  
                           \text{ if } (W) = (\underbrace{i,\dots,i}_{w_i }) \\                
     \end{cases} \; .
  \eea

\qed

\begin{lemma}\label{lem:lema5}
  We have $(\cT_2 \star_{(\;)} \cT_1)^i \sim (\cT_1 \star_{(\;)} \cT_2)^i $.
\end{lemma}
{\bf Proof:}
We have
\bea
 \cT_2 \star_{(\;)} \cT_1 = \begin{cases}
                             (k,Y_1) &\text{ for all } (k,Y_1)\in \cT_1 \\
                             ( \underbrace{i,\dots i }_{n_i+1}, Y_2 ) &\text{ for all } (i,Y_2) \in \cT_2
                            \end{cases} \; ,
\eea
hence
\bea\label{eq:branchesi}
(\cT_2 \star_{(\;)} \cT_1)^i = \begin{cases}
                               (Y_1) &\text{ for all } (i,Y_1)\in \cT_1 \\
                               ( \underbrace{i,\dots i }_{n_i}, Y_2 ) &\text{ for all } (i,Y_2) \in \cT_2
                            \end{cases}
                            =  \begin{cases}
                               (Y_1) &\text{ for all } Y_1\in \cT_1^i \\
                               ( \underbrace{i,\dots i }_{n_i}, Y_2 ) &\text{ for all } Y_2 \in \cT_2^i
                            \end{cases} \; .
\eea
Note that the vertices $(Y_1)\in \cT_1^i$ start necessarily by at most $n_i-1$ labels $i$.
The maximal vertex of color $i$ in $ (\cT_2 \star_{(\;)} \cT_1)^i$ is $(\underbrace{i,\dots,i}_{n_i+q_i-1})$.
The tree $ (\cT_2 \star_{(\;)} \cT_1)^i  $ is equivalent with $\cT'$ obtained by mapping the 
vertices
\bea
 (i,V) \in  (\cT_2 \star_{(\;)} \cT_1)^i \rightarrow (V) \in \cT' 
\qquad (W)\in \cT, (W)\neq(i,V) \rightarrow ( \underbrace{i,\dots,i}_{n_i+q_i-1} ,W) \;.
\eea 
Iterating $n_i$ times we see that all the vertices belonging initially to $\cT_1^i$ will, at one step,
acquire $n_i+q_i-1 $ labels $i$ and lose a label $i$ at all the other $n_i-1$ steps. The vertices belonging to
$\cT_2$ will only lose a label $i$ at all $n_i$ steps. Thus $(\cT_2 \star_{(\;)} \cT_1)^i   $ is equivalent
to the tree
\bea
 \begin{cases}
    ( \underbrace{i,\dots,i}_{q_i} ,Y_1) &\text{ for all } Y_1\in \cT_1^i \\
    (Y_2) &\text{ for all } Y_2 \in \cT_2^i
 \end{cases} \; ,
\eea
which we recognize by eq. \eqref{eq:branchesi} to be $ (\cT_1 \star_{(\;)} \cT_2)^i  $.

\qed

\subsection{The Lie algebra indexed by colored rooted D-ary trees}

We now define a Lie algebra of operators indexed by the trees.  We associate to every tree a variable $t_{\cT}$,
and we denote $|R_1|$ the coordination of the root of $\cT_1$. Consider the operators $\cL_{\cT_1}$ defined as 
\bea\label{eq:defcl}
 \cL_{\cT_1} &=&  (-)^{|R_1|}N^{D- D |R_1|} \;  
  \frac{\partial^{|R_1|}} {
\prod_{i, \cT_1^i \neq \emptyset }
\partial t_{\cT_1^i}}  
+ \sum_{\cT} t_{\cT} \sum_{V \in \cT }  \frac{\partial}{\partial t_{\cT \star_V \cT_1}} \; ,\crcr
\cL_{ \{(\;) \} }&=&  N^D +  \sum_{\cT} t_{\cT} \sum_{V \in \cT }  \frac{\partial}{\partial t_{\cT} } \; ,
\eea
where $N$ is some parameter (destined to become the large $N$ parameter of the tensor model).
We will not consider the most general domain of this operators. Namely, in stead of defining them for arbitrary functions
$f(t_{\cT})$ we restrict their domain to class functions $f(t_{[\cT]})$, with $t_{[\cT]} = \sum_{\cT'\sim \cT} t_{\cT'}$.

\begin{theorem}\label{th:comutator}
 When restricted to class functions, the operators $\cL_{\cT}$ form a Lie algebra with commutator
\bea
 \Bigl[ \cL_{\cT_2} , \cL_{\cT_1} \Bigr] f(t_{[\cT]})  && = \sum_{V \in \cT_2 } \cL_{ \cT_2 \star_V \cT_1 }f(t_{[\cT]}) -
  \sum_{V \in \cT_1} \cL_{ \cT_1 \star_V \cT_2 }f(t_{[\cT]}) \; .
\eea
\end{theorem}
{\bf Proof:} We start by evaluating $\cL_{\cT_2} \cL_{\cT_1}f$. We have
\bea
 \cL_{\cT_2} \cL_{\cT_1} f =&& 
\Big{[} (-)^{|R_2|} N^{D-|R_2|D} \;  
 \frac{\partial^{|R_2|} }{ \prod \partial t_{\cT_2^j}} 
+ \sum_{\cT'} t_{\cT'}  \sum_{V' \in \cT' }   \frac{\partial}{\partial t_{\cT' \star_{V'} \cT_2}} \
\Big{]} \\
&& \Big{[}  (-)^{|R_1|} N^{D-|R_1|D} \; 
 \frac{\partial^{|R_1|}f }{ \prod \partial t_{\cT_1^i}} 
+ \sum_{\cT} t_{\cT} \sum_{V \in \cT }  \frac{\partial f}{\partial t_{\cT \star_V \cT_1}}  
\Big{]}\crcr
&& = (-)^{|R_1|+|R_2|} N^{2D-D(|R_1|+|R_2|)} \frac{\partial^{|R_1|+|R_2|} f }{ \prod \partial t_{\cT_{2}^j} 
\prod \partial t_{\cT_{1}^i} } \crcr
&&+ (-)^{|R_2|} N^{D-|R_2|D}  \Big{(} \sum_{k } \sum_{V\in \cT_2^k}
 \frac{\partial^{|R_2|} f }{ \prod_{j \neq k}\partial t_{\cT_{2}^j} \partial t_{\cT_2^k \star_V \cT_1}} 
+ \sum_{\cT} t_{\cT} \sum_{V \in \cT }  
\frac{\partial^{|R_2|+1} f}{ \prod_{j} \partial t_{\cT_2^{j}} \partial t_{\cT \star_V \cT_1}} \Big{)}\crcr
&&+ (-)^{|R_1|} N^{D-|R_1|D} \sum_{\cT'} t_{\cT'}  \sum_{V' \in \cT' }
 \frac{\partial^{|R_1|+1}f}{ \prod \partial t_{\cT_1^i} \partial t_{\cT'\star_{V'} \cT_2}} \crcr
&&+ \sum_{\cT'} t_{\cT'} \sum_{V'\in\cT'} \Big{(} \sum_{V \in \cT' \star_{V'} \cT_2 }
 \frac{\partial f}{\partial t_{ (\cT' \star_{V'} \cT_2  ) \star_V \cT_1  } } + 
 \sum_{\cT}  t_{\cT} \sum_{ V \in \cT} 
\frac{\partial^2 f}{ \partial t_{\cT'\star_{V'} \cT_2} \partial t_{\cT \star_V \cT_1}} \Big{)} \; ,
\nonumber
\eea
hence the commutator is
\bea
 \Bigl[ \cL_{\cT_2} , \cL_{\cT_1} \Bigr] f = && (-)^{|R_2|} N^{D-|R_2|D} \sum_{k } \sum_{V\in \cT_2^k}
 \frac{\partial^{|R_2|} f }{ \prod_{j \neq k}\partial t_{\cT_{2}^j} \partial t_{\cT_2^k \star_V \cT_1}} 
\\
&& -(-)^{|R_1|} N^{D-|R_1|D} \sum_{k } \sum_{V\in \cT_1^k}
 \frac{\partial^{|R_1|} f }{ \prod_{j \neq k}\partial t_{\cT_{1}^j} \partial t_{\cT_1^k \star_V \cT_2}}  \crcr
&&+ \sum_{\cT} t_{\cT} \sum_{V'\in\cT} \sum_{V \in \cT \star_{V'} \cT_2 }
 \frac{\partial f}{\partial t_{ (\cT \star_{V'} \cT_2  ) \star_V \cT_1  } }
-\sum_{\cT} t_{\cT} \sum_{V'\in\cT} \sum_{V \in \cT \star_{V'} \cT_1 }
 \frac{\partial f}{\partial t_{ (\cT \star_{V'} \cT_1  ) \star_V \cT_2  } }  \; . \nonumber
\eea

We reorganize the terms.  For the terms in the first line we use lemma \ref{lem:lema1}
and identify the derivatives with respect to $(\cT_2 \star_V \cT_1)^k  $ (resp. 
$(\cT_1 \star_V \cT_2)^k$) .
For the terms in the second line, the vertex $V$ can either be the image of a vertex in  
$\cT$ or of a vertex in $\cT_2$ (resp. $\cT_1$). Separating the term with 
$V$ the image of $V' \in \cT$ we get 
\bea
 \Bigl[ \cL_{\cT_2} , \cL_{\cT_1} \Bigr] f = && (-)^{|R_2|} N^{D-|R_2|D} \sum_{V \in \cT_2 \setminus (\;)}
\frac{\partial^{|R_2|} f }{ \prod_{j }\partial t_{ (\cT_2 \star_V \cT_1)^j   }} \\
&&- (-)^{|R_1|} N^{D-|R_1|D} \sum_{V \in \cT_1 \setminus (\;) }
\frac{\partial^{|R_1|} f }{ \prod_{j }\partial t_{ (\cT_1 \star_V \cT_2)^j   }} \crcr
&&+\sum_{\cT} t_{\cT} \sum_{\stackrel{V, V'\in\cT}{ V \neq V'}} 
\Big{(} \frac{\partial f}{\partial t_{ (\cT \star_{V'} \cT_2  ) \star_V \cT_1  } }
- \frac{\partial f}{\partial t_{ (\cT \star_{V'} \cT_1  ) \star_V \cT_2  } } \Big{)}\crcr
&&+ \sum_{\cT} t_{\cT} \sum_{V'\in\cT} \Big{(} \frac{\partial f}{\partial t_{ (\cT \star_{V'} \cT_2  ) \star_{V'} \cT_1  } }
- \frac{\partial f}{\partial t_{ (\cT \star_{V'} \cT_1  ) \star_{V'} \cT_2  } } 
\Big{)} \crcr
&&+  \sum_{\cT} t_{\cT} \sum_{V'\in\cT} \sum_{V \in \cT_2 \setminus (\;)}
 \frac{\partial f}{\partial t_{ (\cT \star_{V'} \cT_2  ) \star_V \cT_1  } }
-\sum_{\cT} t_{\cT} \sum_{V'\in\cT} \sum_{V \in \cT_1 \setminus (\;) }
 \frac{\partial f}{\partial t_{ (\cT \star_{V'} \cT_1  ) \star_V \cT_2  } } \; , \nonumber
\eea 
where, by a slight abuse of notations, we denote $V$ and $V'$ also the images of $V$ and $V'$ under
$\star$ operations.
By lemma \ref{lem:lema3}, the terms in the third line cancel (after exchanging $V$ and $V'$ in the second term).
Using lemma \ref{lem:lema4} the terms in the last line recombine with the ones in the first two lines.
Finally the terms in the fourth line rewrite using lemma \ref{lem:lema2}. We thus obtain 
\bea
 \Bigl[ \cL_{\cT_2} , \cL_{\cT_1} \Bigr] f && = \sum_{V \in \cT_2 \setminus (\;)} \cL_{ \cT_2 \star_V \cT_1 }f -
  \sum_{V \in \cT_1 \setminus (\;) } \cL_{ \cT_1 \star_V \cT_2 }f \crcr
&&+ \sum_{\cT} t_{\cT} \sum_{V \in \cT} \Big{(} 
 \frac{\partial f} {\partial t_{\cT \star_{V} ( \cT_2 \star_{ (\;) } \cT_1 )}  }
- \frac{\partial f} {\partial t_{\cT \star_{V} ( \cT_1 \star_{ (\;) } \cT_2 )}  }
   \Big{)} \; .
\eea

Note that in both trees $ \cT_1 \star_{ (\;) } \cT_2 $ and 
$\cT_2 \star_{ (\;) } \cT_1 $, the root has a nonempty branch of color $i$ if at least one of
$\cT_1^i$ or $\cT_2^i$ is non empty. The two roots have then equal coordination denoted
$|R_{12}|$. Adding and subtracting, the commutator becomes 
\bea
\Bigl[ \cL_{\cT_2} , \cL_{\cT_1} \Bigr] f && = \sum_{V \in \cT_2 } \cL_{ \cT_2 \star_V \cT_1 }f -
  \sum_{V \in \cT_1} \cL_{ \cT_1 \star_V \cT_2 }f 
 - N^{D - D|R_{12}|} \Big{(} \frac{\partial^{|R_{12}|}f }{\prod t_{ (\cT_2 \star_{(\;)} \cT_1)^i}} 
 - \frac{\partial^{|R_{12}|}f }{\prod t_{ (\cT_1 \star_{(\;)} \cT_2)^i}} 
\Big{)}  \; ,
\eea
and the last term cancels due to lemma \ref{lem:lema5} and taking into account that $f$ is a class 
function $f([\cT])$, thus $\partial_{t_{\cT}}f = \partial_{t_{\cT'}}f$ if $\cT\sim \cT'$.

\qed

\section{Schwinger Dyson equations in the large N limit of Colored Tensor Models}\label{sec:SDE}

In this section we first recall the independent identically distributed (i.i.d.) colored tensor model
and its $1/N$ expansion. We then generalize it to a colored model with a generic potential, derive,
in the large $N$ limit the SDEs of the model and translate them into a set of equations 
(involving the operators $\cL_{\cT_1}$) for the partition function $Z$.
We will closely parallel the derivation of the loop equations in 
section \ref{sec:matrixmodels}. 

\subsection{The i.i.d. colored tensor models with one coupling}

We denoted $\vec n_i$, for $i=0,\dotsc,D$, the $D$-tuple of integers
 $\vec n_i = (n_{ii-1},\dotsc, n_{i0},\; n_{iD}, \dotsc, n_{ii+1}) $, with
$n_{ik}=1,\dotsc, N$. This $N$ is the size of the tensors and the large $N$ limit defined in 
\cite{Gur3,GurRiv,Gur4} represents the limit of infinite size tensors. 
We set $n_{ij} = n_{ji}$. Let $\bar \psi^i_{\vec n_i},\; \psi^i_{\vec n_i}$, with $i=0,\dotsc, D$, be $D+1$ couples of complex
conjugated tensors with $D$ indices. The independent identically distributed (i.i.d.) colored tensor
model in dimension $D$ \cite{color,lost,Gur4,PolyColor} is defined by the partition function
\begin{gather}
\nonumber e^{ - N^D F_N(\lambda,\bar\lambda)} = Z_N(\lambda, \bar{\lambda}) = \int \, d\bar \psi \, d \psi
\ e^{- S (\psi,\bar\psi)} \; , \\
S (\psi,\bar\psi) = \sum_{i=0}^{D} \sum_{ n} \bar \psi^i_{\vec n_i} \psi^i_{\vec n_i}  +
\frac{\lambda}{ N^{D(D-1)/4} } \sum_{ n} \prod_{i=0}^D \psi^i_{ \vec n_i } +
\frac{\bar \lambda}{ N^{D(D-1)/4} } \sum_{ n}
\prod_{i=0}^D \bar \psi^i_{ \vec n_i } \; . \label{eq:iid}
\end{gather}
$\sum_{ n}$ denotes the sum over all indices $ n_{ij}$ from $1$ to $N$.
The tensor indices $n_{ij}$ need not be simple integers (they can for instance index the
Fourier modes of an arbitrary compact Lie group, or even of a finite group of large 
order \cite{bianca-finitegp}). Rescaling $\psi^i_{\vec n_i} = N^{-D/4} P^i_{\vec n_i}$ leads to
\bea
 S (\bar P ,P  ) = N^{D/2} \Big{(}
\sum_{i=0}^{D} \sum_{\vec n} \bar P^i_{\vec n_i} P^i_{\vec n_i}  +
\lambda \sum_{\vec n} \prod_{i=0}^D P^i_{ \vec n_i } +
\bar \lambda \sum_{\vec n} \prod_{i=0}^D \bar P^i_{ \vec n_i } \Big{)} \;.
\eea
The partition function of equation \eqref{eq:iid} is evaluated by colored stranded Feynman graphs
\cite{color,lost,PolyColor}. The tensors have {\it no} symmetry properties under permutations 
of their indices (i.e. all $\psi^i_{\vec n_i}, \bar \psi^i_{\vec n_i}$ are independent).
The colors $i$ of the fields $\psi^i, \bar{\psi}^i$ induce important
restrictions on the combinatorics of stranded graphs. We have two types of vertices,
say one of positive (involving $\psi$) and one of negative (involving $\bar \psi$).
The lines always join a $\psi^i$ to a $\bar{\psi}^i$ and possess a color index.
Any Feynman graphs $\cG$ of this model is a simplicial pseudo manifold \cite{lost} and
the colored tensor models provide a statistical theory of random triangulations in 
dimensions $D$, generalizing random matrix models.
The tensor indices $n_{jk}$ are preserved along the strands. The amplitude of a graph with $2p$ 
vertices and $\cF$ faces (closed strands) is \cite{Gur4}
\bea\label{eq:ampli}
A(\cG)= (\lambda\bar\lambda)^p N^{-p \frac{D(D-1)}{2}+ \cF } \; .
\eea

The {\bf $n$-bubbles} of the graph  are the maximally connected subgraphs made of lines with $n$ fixed colors.
For instance, the $D$-bubbles are the maximally connected subgraphs containing
all but one of the colors. 
They are associated to the $0$ simplices (vertices) of the pseudo-manifold.
We label $\cB^{\widehat{i}}_{(\rho)}$
the $D$-bubbles with colors $\{0,\dots, D\} \setminus \{i \}$ (and $\rho$ labels the various bubbles
with identical colors).
We denote $\cB^{[D]}$ the total number of $D$ bubbles, which respects
\cite{Gur4}
\bea\label{eq:ineqpDBD}
 p + D - \cB^{[D]} \ge 0 \;,
\eea
where $p$ is half the number of vertices of the graph.

A second class of graphs crucial for the $1/N$ expansion of the colored tensor model are the
{\bf jackets} \cite{sefu3,Gur3,GurRiv,Gur4}.
\begin{definition}
Let $\tau$ be a cycle on $\{0,\dotsc,D\}$. A colored {\bf jacket} $\cJ$ of $\cG$ is the ribbon graph made by faces
with colors $(\tau^q(0),\tau^{q+1}(0))$, for $q=0,\dotsc,D$, modulo the orientation of the cycle.
\end{definition}
A jacket $\cJ$ of $\cG$ contains all the vertices and all the lines of $\cG$ (hence $\cJ$ and $\cG$ have the same
connectivity), but only a subset of faces. The
jackets (further studied in \cite{BS3,Ryan:2011qm}) are ribbon graphs, completely
classified by their genus $g_\cJ$. For a colored graph $\cG$ we define its {\it degree} \cite{GurRiv,Gur4}
\begin{definition}
The {\bf degree} $\omega(\cG)$ of a graph is the sum of genera of its jackets, $\omega(\cG)=\sum_{\cJ} g_{\cJ}$.
\end{definition}

The number of faces of a graph evaluates as a function of its degree \cite{GurRiv,Gur4}
\bea\label{eq:propdegree}
  &&  \omega(\cG) = \frac{(D-1)!}{2} \Big{(} p+D-\cB^{[D]} \Big{)} + \sum_{i;\rho} \omega(\cB^{\widehat{i}}_{(\rho)}) \; .
   \crcr
  && \frac{2}{(D-1)!} \omega(\cG) = \frac{D (D-1)}{2} p + D - \cF \; .
\eea

The $1/N$ expansion of the colored tensor model is encoded in the remark that $\omega(\cG)$, which is 
a positive number, has exactly
the combination of $p$ and $\cF$ appearing in the amplitude of a graph \eqref{eq:ampli}, thus
\bea
A(\cG) = (\lambda\bar\lambda)^p\ N^{D - \frac{2}{(D-1)!} \omega(\cG)  }\;.
\eea
The free energy $F_N(\lambda, \bar{\lambda})$ of the model admits then an expansion in the degree
\bea
F_N(\lambda,\bar\lambda) = F_{\infty} ( \lambda,\bar\lambda ) + O(N^{-1} ) \;,
\eea
where $F_{\infty}( \lambda,\bar\lambda ) $ is the sum over all graphs of degree $0$.
The degree plays in dimensions $D\geq 3$ the role played by the genus in matrix models, and in
particular degree $0$ graphs are spheres \cite{GurRiv}.

\begin{lemma} \label{planar jacket}
If the degree vanishes (i.e. all jackets of $\cG$ are planar) then $\cG$ is dual to a $D$-sphere.
\end{lemma}

We conclude this section with the following lemma.

\begin{lemma}\label{lem:ddegg}
Let $\cG$ be a graph (with colors $0,\dots D$) and $\cB^{ \widehat{D} }_{(\rho)}$ its $D$-bubbles
with colors $0,\dots, D-1$. Then
\bea
 \omega(\cG) \ge D \sum_{\rho} \omega (\cB^{\widehat{D} }_{(\rho)} ) \; .
\eea
\end{lemma}
{\bf Proof:}
Consider a jacket $\cJ$ of $\cG$. By eliminating the color $D$ in its associated cycle we obtain 
a cycle over $0,\dots,D-1$ associated to a jacket $\cJ^{\widehat{D}}_{(\rho)}$ for each of its bubbles.
As graphs, $ \cJ^{\widehat{D}}_{(\rho)} $ are one to one with disjoint subgraphs of $\cJ$
(obtained by deleting the lines of color $D$ and joining the strands $(\pi^{-1}(D),D)$
and $(D,\pi(D))$ in mixed faces corresponding to $(\pi^{-1}(D),\pi(D))$
in $ \cJ^{\widehat{D}}_{(\rho)} $ \cite{GurRiv}), consequently
\bea
 g_{\cJ} \ge \sum_{(\rho) } g_{\cJ^{\widehat{D}}_{(\rho)}  } \; .
\eea
Every jacket $ \cJ^{\widehat{D}}_{(\rho)} $ is obtained as subgraph of exactly $D$ distinct jackets 
$\cJ$ (corresponding to inserting the color $D$ anywhere in the cycle associated to 
$ \cJ^{\widehat{D}}_{(\rho)} $ $ ) $. Summing over all jackets of $\cG$ we obtain
\bea
 \sum_{J} g_{\cJ} \ge D \sum_{\rho } \sum_{ \cJ^{\widehat{D} }_{ (\rho) } } g_{\cJ^{\widehat{D}}_{(\rho)}  } \; .
\eea

\qed
 
\subsection{Leading order graphs}\label{sec:melons}

The leading order graphs of the colored tensor model have been analyzed in detail in \cite{Bonzom:2011zz}. 
We present below reader's digest of these results. We are interested in understanding in more depth the 
structure of leading order vacuum graphs in $D$ dimensions. Leading order vacuum graph can be obtained from 
leading order two point graphs by reconnecting the two external lines (and conversely, cutting any line in a 
leading order vacuum graph leads to a leading order two point graph). We detail below the two point graphs.

\begin{figure}[htb] 
\begin{center}
 \includegraphics[width=5cm]{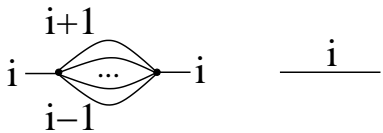}
 \caption{ Eliminating a $D$-bubble with two vertices. }
\label{fig:redmelon}
 \end{center}
\end{figure}

A $D$-bubble with two vertices
$\cB^{\widehat{i}}_{(\rho)}$ has $\frac{D(D-1)}{2}$ internal faces, hence, by equation
\eqref{eq:propdegree}, the degree (and the topology) of a graph $\cG$ and of the graph
$\cG_{/\cB^{\widehat{i}}_{(\rho)}}$  obtained by replacing $\cB^{\widehat{i}}_{(\rho)} $
with a line of color $i$ (see figure \ref{fig:redmelon}) are
identical\footnote{This elimination is a $1$-Dipole contraction for one of the two
lines of color $i$ touching $\cB^{\widehat{i} }_{(\rho)}$ \cite{Gur4}.}.

It can be shown \cite{Bonzom:2011zz} that, for $D\ge 3$, a leading order two point graph must possess a $D$-bubble with exactly 
two vertices. Eliminating this bubble, we obtain a leading order graph having two less vertices. The new graph must 
in turn possess a bubble with two vertices, which we eliminate, and so on. It follows that the leading order 
2-point graphs must reduce after a sequence of eliminations of $D$-bubbles to the graph with a single $D$-bubble
and only two vertices of figure \ref{fig:redmelon}. It is more useful to take the reversed point of view and start with the 
graph of figure \ref{fig:redmelon} and insert $D$-bubbles with two vertices on its lines.
This insertion procedure preserves colorability, degree and topology.
The leading order 2-point connected graphs (with external legs of color say $D$) are in one to one correspondence 
with colored rooted $(D+1)$-ary trees.

\begin{figure}[htb]
 \includegraphics[width=5cm]{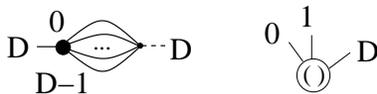}
 \caption{First order.}
  \label{fig:order1}
\end{figure}

{\bf Order $(\lambda\bar \lambda)$:} The lowest order graph consists in exactly one $D$-bubble with two vertices
(and external lines say of color $D$). We represent this graph by the tree with only the root vertex $(\;)$
decorated with $(D+1)$ leaves. The leaves have colors $0,\dots D$. On $\cG$, we consider 
``active'' all lines of colors $j\neq D$ and the line of color $D$ touching the vertex $\lambda$. 
They correspond to the leaves of the vertex. See figure \ref{fig:order1},
where the vertex $\lambda$ is dotted and the inactive line is represented as dashed.

\begin{figure}[htb]
 \includegraphics[width=5cm]{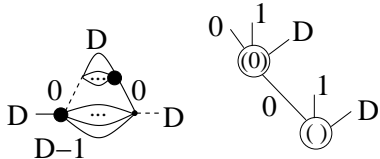}
 \caption{Second order.}
  \label{fig:order2}
\end{figure}

{\bf Order $(\lambda \bar \lambda)^2$:} At second order we have $D+1$ graphs contributing.
They come from inserting a $D$-bubble with two vertices on any of the $D+1$ active lines of the
first order graph. All the interior lines of the new $D$-bubble are active, and so is the exterior
line touching its vertex $\lambda$. Say we insert the new bubble on the active line of color 
$j$. This graph corresponds to the tree $\{(\;), (j)\}$, see figure \ref{fig:order2} for the case $j=0$.

{\bf Order $(\lambda \bar\lambda)^{p+1} $:} We obtain the graphs at order $p+1$ by inserting a $D$-bubble
with two vertices on any of the active lines of a graph at order $p$. The interior lines (and the exterior line
touching the vertex $\lambda$) of the new bubble are active. We represent this by connecting a vertex of coordination
$D+2$, with $D+1$ active leaves, on one of the active leaves of a tree at order $p$. The new tree line inherits the color
of the active line on which we inserted the $D$-bubble. At order $(\lambda \bar\lambda)^p$ we obtain contributions from all
rooted colored $(D+1)$-ary trees with $p$ vertices. 

Our tree is a colored version of Gallavotti-Nicolo \cite{Gal} tree.
The vertices of the tree represent certain subgraphs of $\cG$. We call them {\bf melons} $\cM$ 
and we identify them as the 1-particle irreducible (1PI) amputated 2-point 
sub-graphs of $\cG$. The intuitive picture is that a
melon is itself made of melons within melons. The $D$-bubbles with only
two vertices are obviously the smallest melons. The largest melon is the graph itself.

A rooted tree is canonically associated to a partial order. The partial ordering corresponding
to the tree we have introduced is
\bea \label{ordered melons}
 \cM_1 \geq \cM_2 \quad \text{ if } \ \begin{cases}
         \text{either}\quad \cM_1 \supset \cM_2 \\
         \text{or} \quad \left\{ \begin{aligned}
         &\exists \;  \cN_{(\rho)},
         \; \cM_1 \cup \bigl( \cup_{\rho} \cN_{(\rho)} \bigr) \cup \cM_2
          \text{ is a 2-point }\\
         & \text{  amputated connected sub-graph of $\cG$ }\\
         & \text{ with external points } \bar\lambda \in \cM_1 \text{ and }
         \lambda \in \cM_2 \end{aligned}\right.
                          \end{cases} \; ,
\eea
and $\ge$ is transitive.

\begin{figure}[htb]
\begin{center}
 \includegraphics[width=10cm]{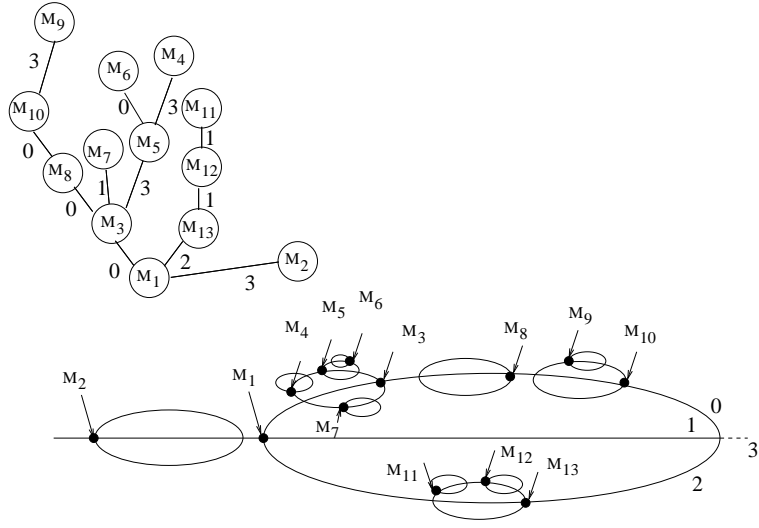}
\caption{A melon graph and its associated colored GN rooted tree.}
\label{fig:t5}
\end{center}
\end{figure}

The line connecting $\cM$ towards the root on the tree (i.e. going to a greater melon) inherits the
color of the exterior half-lines of $\cM$. An example in $D=3$ is given in
figure \ref{fig:t5} where the dotted vertices of $\cG$ are $\lambda$, the inactive line 
of $\cG$ is dashed and the active leaves are implicit. We identify the melons
by their external point $\lambda$. Since the active external line of a melon is always chosen to be
the one touching the vertex $\lambda$, the root melon in an arbitrary graph is the one containing the external
point $\bar\lambda$, e.g. $\cM_1$ in figure \ref{fig:t5}. Note that $\cM_3\supset \cM_4,\cM_5,\cM_6,\cM_7$,
hence it is their ancestor. Also $\cM_3 \cup \cM_8 \cup \cM_{10}$ forms a two point function with external point
$\bar\lambda \in \cM_3 $ and as $\cM_9 \subset \cM_{10}$, the melon $\cM_3$ is the ancestor of
$ \cM_4,\cM_5,\cM_6,\cM_7, \cM_8,\cM_9,\cM_{10}$.

The vacuum leading order graphs (also called melonic) are obtained by reconnecting the two exterior half lines of a 
melonic two point graphs with a line. Their amplitude is $N^D$. If a graph is a melonic graph with $D+1$ colors, all
its $D$ bubbles are melonic graphs with $D$ colors. This is easy to see, as the reduction of a $D$ bubble with 
two vertices represents the reduction of a $D-1$ bubble with two vertices for all $D$ bubbles which contain 
the two vertices. When reducing the graph to its root melon, one by one all its $D$-bubbles reduce to
$D$-bubbles with two vertices. 

Moreover, the $D$-ary trees of the $D$-bubbles are trivially obtained from the
tree of the graph by deleting all lines (and leaves) of color $D$. 
We will be needing below the following obvious fact: given a melonic graph and one of its
$D$ bubbles (say $\cB^{\widehat{D}}_{(1)}$), all the lines of color $D$ connecting 
on it either separate it from a different $D$-bubble (and are tree lines in the associated 
colored rooted tree) or they connect the two external points
of a 1PI amputated two point subgraph with $D-1$ colors of $\cB^{\widehat{D}}_{(1)}$
(i.e. they connect the two external points of a melon in $\cB^{\widehat{D}}_{(1)}$), in which
case they are leaves of the associated tree.

\subsection{From one to an infinity of coupling constants}

Inspired by section \ref{sec:matrixmodels}, we generalize the colored tensor model with one coupling to a model 
with an infinity of couplings and derive the SDEs of the general model. 
First we integrate all colors save one, and second we ``free'' the couplings of the operators 
in the effective action for the last color.

When integrating all colors save one the partition function becomes
\bea
&& Z = \int d\psi^D d\bar \psi^D \; e^{-S^D(\psi^D, \bar \psi^D)} \crcr
&& S^D(\psi^D, \bar \psi^D) =  \sum \bar \psi^D_{ \vec n_D } \psi^D_{\vec n_D }
 + \sum_{ \cB^{\widehat{D}} } \frac{(\lambda\bar\lambda)^{p} }{ \text{Sym}(\cB^{\widehat{D}}) } 
\; \tr_{\cB^{\widehat{D}}} [\bar\psi^D, \psi^D ]
\;  N^{-\frac{D(D-1)}{2}p + \cF_{\cB^{\widehat{D}} } }
\eea
where the sum over $\cB^{\widehat{D}}$ runs over all connected vacuum graphs with colors $0,\dots D-1$ 
(i.e. over all the possible $D$-bubbles with colors $0,\dots D-1$) and $p$ vertices.
The operators $\text{Tr}_{\cB^{\widehat{D}} } [\bar\psi^D, \psi^D ] $ in the effective action for the last color
are tensor network operators. Every vertex of $\cB^{\widehat{D}} $ is decorated by a tensor
$\psi^D_{\vec n_D}$ or $\bar \psi^D_{ \vec {\bar n}_D }$, and the tensor indices are contracted 
as dictated by the graph $\cB^{\widehat{D}} $. 
We denote $v$, $\bar v$ the positive (resp. negative) vertices of $\cB^{\widehat{D}} $, 
and $l^i_{v\bar v}$ the lines (of color $i$)
connecting the positive vertex $v$ with the negative vertex $\bar v$.
The operators write
\bea\label{eq:tensnet}
 \text{Tr}_{\cB^{\widehat{D}}} [\bar\psi^D, \psi^D ] = \sum_{n}
 \Bigl( \prod_{v,\bar v \in \cB^{\widehat{D}}} \bar \psi^D_{\vec {\bar  n}^{\bar v}_D } \psi^D _{\vec n^v_D } \Bigr)
 \Bigl( \prod_{i=0}^{D-1} \prod_{ l^i_{v\bar v} \in \cB^{\widehat{D}}  } 
\delta_{n^{v }_{Di} {\bar n}^{\bar v }_{Di} } \Bigr) \; ,
\eea
where all indices $n$ are summed. Note that, as all vertices in the bubble belong to an unique line of a given color, all
the indices of the tensors are paired. The scaling with $N$ of an operator computes in terms of its degree
\bea
 N^{-\frac{D(D-1)}{2}p + \frac{(D-1)(D-2)}{2} p + D-1-\frac{2}{(D-2)!}\omega(\cB^{\widehat D}) } 
= N^{ - (D-1) p + D-1 -\frac{2}{(D-2)!}\omega(\cB^{\widehat D}) } \;,
\eea
thus the effective action for the last color writes (dropping the index $D$)
\bea
 S^D ( \psi , \bar \psi) =  \sum \bar \psi_{ \vec  n } \; \psi_{  \vec n }
 + N^{D-1} \sum_{ \cB }  \frac{ (\lambda \bar\lambda)^{ p  } }{ \text{Sym}( \cB  )} 
\; N^{-(D-1)p -\frac{2}{(D-2)!}\omega(\cB) } \;
\text{Tr}_{ \cB  }  [ \bar\psi  , \psi ] \; ,
\eea
Attributing to each operator its coupling constant and rescaling the field to $T = \psi N^{-\frac{D-1}{2}}$, we
obtain the partition function of colored tensor model with generic potential 
\bea\label{eq:genmod}
&& Z =e^{-N^D F(g_{\cB } ) }= \int d\bar T dT \; e^{-N^{D-1} S(\bar T, T)} \; ,\crcr
&& S (\bar T, T ) = \sum \bar T_{\vec n} T_{\vec n} +   \sum_{ \cB  }  t_{ \cB  } \quad  
N^{ -\frac{2}{(D-2)!}\omega(\cB) } \; \text{Tr}_{ \cB } [ \bar T ,  T ] \; .
\eea

It is worth noting that, although in the end we deal with an unique tensor $T$, the colors are crucial to the
definition of the tensor network operators in the effective action. The initial vertex of the tensor model
described a $D$ simplex. The tensor network operators describe (colored) polytopes in $D$ dimensions obtained by 
gluing simplices along all save one of their faces around a point (dual to the bubble $\cB $).
This is in strict parallel with matrix models, where higher degree interactions represent polygons obtained by
gluing triangles around a vertex.

When evaluating amplitudes of graphs obtained by integrating the last tensor $T$, the tensor network operators act 
as effective vertices (for instance each comes with its own coupling constant). It is however more convenient 
to represent the Feynman graph of the path integral \eqref{eq:genmod} still as graphs with $D+1$ colors. The
effective vertices are the subgraphs with colors $0,\dots,D-1$, and encode the connectivity of the tensor 
network operators. 

The partition function of eq. \eqref{eq:genmod} provides a natural set of observables of the model: the multi 
bubble correlations defined as 
\bea
 \Big\langle  \text{Tr}_{\cB_{(1)} } [\bar T, T] \; \text{Tr}_{\cB_{(2)}  } [\bar T, T] 
\dots \text{Tr}_{\cB_{(\rho )} } [\bar T, T] \Big\rangle 
 =  \prod_{i=1}^{\rho} \Big{(} -N^{- \Bigl[D-1- \frac{2}{(D-2)!}\omega(\cB_{(i) }) \Bigr]}
\frac{\partial} {\partial t_{ \cB_{(i)} } } \Big{)}Z \; .
\eea

When introducing an infinity of coupling constants, we did not change the scaling with $N$ of the 
operators. The graphs $\cG$ contributing to the connected multi bubble correlations
are connected vacuum graphs with $D+1$ colors and with $\rho$ marked subgraph corresponding to the insertions
$\tr_{\cB_{(\nu)} } [\bar T, T]$. Taking into account the scaling of the insertions, the global scaling
of such graphs is
\bea\label{eq:scaling}
&&\Big\langle  \text{Tr}_{\cB_{(1)}  } [\bar T, T] \; \text{Tr}_{\cB_{(2)}   } [\bar T, T] 
\dots \text{Tr}_{\cB_{(\rho )}  } [\bar T, T] \Big\rangle_c \crcr
&& \le  N^{D - \frac{2}{(D-1)!} \omega(\cG)} N^{- \rho (D-1) + \sum_{\rho} \frac{2}{(D-2)!} \omega(\cB_{(\rho)})}
\le N^{ D - \rho (D-1)  - \frac{2}{D!} \omega(\cG) }
\; ,
\eea 
where we use lemma \ref{lem:ddegg}. 

In the large $N$ limit, the connected correlations
receiving contributions from graphs of degree $0$ (melonic graphs) dominate the multi
bubble correlations. All their bubbles are necessarily melonic,
in particular the insertions $\text{Tr}_{\cB_{(i)}} [\bar T, T] $. 
As we have seen in section \ref{sec:melons}, the melonic $D$-bubbles (i.e. melonic graphs with $D$ colors)
are one to one with colored rooted $D$-ary trees $\cT$. The tensor network operators, eq. \eqref{eq:tensnet},
of melonic bubbles can be written directly in terms of $\cT$. When building a $D$-bubble 
starting from $\cT$, each time we insert
a melon corresponding to a vertex $V\in \cT$ we bring a $T$ and a $\bar T$ tensor for the two external
points of the melon. We denote the indices of 
$T$ by $\vec n_V$ and the ones of $\bar T$ by $\vec {\bar n}_V$, and we get
\bea
\text{Tr}_{ \cB } [ \bar T ,  T ] \equiv
\text{Tr}_{ \cT} [\bar T, T] = \prod_{V \in \cT } \Bigl( T_{ \vec n_V }
\bar T_{ \vec {\bar n}_{V} }
\prod_{i=0}^{D-1} \delta_{n^i_{V} \bar n^i_{s^i_{ \cT }[V] }  } \Bigr) \;,
\eea
where $s^i_{\cT}$ is exactly the colored successor function defined in section \ref{sec:algebra}.
As this operator depends exclusively of the successor functions, it is an invariant for an equivalence 
class of trees $\cT\sim\cT' \Rightarrow \tr_{ \cT} [\bar T, T] = \tr_{ \cT'} [\bar T, T]$,
hence the action and the partition function depend only on the class 
variables $t_{[\cT]} = \sum_{\cT'\sim \cT} t_{\cT}$. Taking into account that the melonic 
bubbles have degree $0$ (and redefining the coupling of the tree $\cT=\{(\;)\}$), the action writes
\bea
&& S (\bar T, T ) = \sum_{\cT}   t_{\cT } \; 
\tr_{ \cT } [ \bar T ,  T ] + S^{r} (\bar T, T) \; ,
\eea
where $S^{r}$ correspond to non melonic bubbles. 

\subsection{Schwinger Dyson equations}

Consider a melonic bubble corresponding to the tree $\cT_1$ with root $(\;)_1$.
We denote
\bea
\delta^{\cT_1}_{n,\bar n} = \prod_{V\in \cT_1} \prod_{i=0}^{D-1} \delta_{n^i_{ V} \bar n^i_{s^i_{ \cT_1 }[V] } } \; .
\eea
The SDEs are deduced starting from the trivial equality
\bea
\sum_{\vec p ,n} \int  \frac{\delta}{\delta  T_{\vec p } }  \Big{[}  
 T_{  \vec n_{ (\;)_1 } } \delta_{ \vec{\bar n}_{ (\;)_1} \vec p}
\Big{(}  \prod_{  V_1 \in \cT_1 \setminus (\;)_1  } T_{ {\vec n}_{V_1} } 
\bar T_{ \vec {\bar n}_{V_1} } \Big{)} \;
\delta^{\cT_1}_{n,\bar n}
 \; \;
e^{-N^{D-1} S } \Big{]} =0 \; ,
\eea
which computes to 
\bea\label{eq:SDEfull}
 \sum_n \int \Big{\{} && \delta_{\vec n_{(\;)_1} \vec {\bar n}_{(\;)_1}  } 
 \Big{(}  \prod_{ V_1 \in \cT_1 \setminus (\;)_1   }  T_{ {\vec n}_{V_1} } 
\bar T_{ \vec {\bar n}_{V_1} } \Big{)} \;
 \delta^{\cT_1}_{n,\bar n} \crcr
&& + \sum_{V_2\neq (\;)_1}
T_{  \vec n_{(\;)_1} } \bar T_{ \vec {\bar n}_{V_2} }  \delta_{\vec {\bar n}_{(\;)_1} \vec n_{V_2}}
 \Big{(}  \prod_{ V_1 \in \cT_1\setminus \{ (\;)_1,V_2 \}   } T_{ {\vec n}_{V_1} } 
\bar T_{ \vec {\bar n}_{V_1} } \Big{)} \; \delta^{\cT_1}_{n,\bar n}
 \crcr
&& - N^{D-1} \; T_{  \vec n_{ (\;)_1} }  \delta_{ \vec{\bar n}_{ (\;)_1} \vec p} \;
\Big{(}  \prod_{ V_1 \in \cT_1\setminus (\;)_1   }  T_{ {\vec n}_{V_1} } 
\bar T_{ \vec {\bar n}_{V_1} } \Big{)} \;\delta^{\cT_1}_{n,\bar n} \crcr
&& \times
\Big{[} \sum_{\cT} t_{\cT} \sum_{V \in \cT }  
  \bar T_{ \vec{\bar n}_{V} }
\delta_{ \vec n_V  \vec p  } \;
 \Big{(}  \prod_{ V' \in \cT \setminus V  } 
T_{ \vec n_{V'} }  \bar T_{ \vec {\bar n}_{V'}  } \Big{)}  \; \delta^{\cT}_{n,\bar n}
+ \frac{\delta S^r}{\delta T_{\vec p}} \Big{]} \Big{\}} \; e^{-N^{D-1}S } \;.
\eea

The second line in eq. \eqref{eq:SDEfull} represents graphs in which a line of color $D$ on a melonic 
bubble connects the $\bar T$ on the root melon $(\;)$ to a $\bar T$ on a distinct 
melon $V_2$. Hence it can not be a melon (see the end of section \ref{sec:melons}).
The last term represents a melonic bubble connected trough 
a line to a non melonic bubble (coming from $\frac{\delta S^r}{\delta T}$).
Thus it can not be a melon either. Taking into account that we have one line explicit in both graphs,
(hence a factor $N^{-(D-1)}$), and that the scaling of $\tr_{\cT} [T,\bar T]$ is $N^{D-1}$, 
in both cases the correlations scale at most like 
\bea\label{eq:nonmelonbound}
\frac{1}{Z} \Big{\langle} \dots \Big{\rangle} \le N^{D-\frac{2}{(D-1)!} } \; .
\eea 

The first term in eq. \eqref{eq:SDEfull} factors over the branches $\cT_1^i$ of $\cT_1$. We denote 
$(\;)_{1,i}$ the root of the branch $\cT_1^i$.
Recall that, for a non empty branch $\cT^i$, the vertex $  s^i_{\cT}[(\;)_1]= (i)\in \cT$ maps
on the root $ (\;)_{1,i} \in \cT^i $. For each branch we evaluate  
\bea
&& \sum_{ n^i_{R_1}, {\bar n}^i_{ (\;)_1}  } \delta_{ n^i_{ (\;)_1} {\bar n}^i_{ (\;)_1}  } 
\delta_{n^i_{ (\;)_1} \bar n^i_{s^i_{\cT_1}[ (\;)_1 ]} }
 \delta_{ n^i_{ [s^i_{\cT}]^{-1}[ (\;)_1]} \bar n^i_{ (\;)_1} } \crcr
&& = 
\begin{cases}
   N & \text{ if } s^i_{\cT}[ (\;)_1]= (\;)_1 \\
\delta_{ n^i_{ [s^i_{\cT}]^{-1}[ (\;)_1]} \bar n^i_{s^i_{\cT_1}[ (\;)_1 ]}   } = 
\delta_{ n^i_{ [s^i_{\cT^i}]^{-1}[ (\;)_{1,i} ] }  \bar n^i_{ (\;)_{1,i} } } & \text{ if not }
\end{cases} \; ,
\eea
thus, denoting $|R_1|$ the coordination of the root $(\;)_1\in \cT_1$, we get
\bea
 \sum_{\vec n_{R_1}, \vec{\bar} n_{R_1}} \delta_{\vec n_{R_1} \vec {\bar n}_{R_1}  } \delta^{\cT_1}_{n,\bar n}
= N^{D-|R_1|} \prod_{ \stackrel{i=0}{\cT^i_1\neq\emptyset }}^{D-1} \delta^{\cT_1^i}_{n,\bar n} \;.
\eea
The third term in eq. \eqref{eq:SDEfull} computes 
\bea
\sum_{ \vec{\bar n}_{ (\;)_1}  \vec n_V } \delta_{\vec{\bar n}_{ (\;)_1}  \vec n_V } 
\delta^{\cT_1}_{n, \bar n} \delta^{\cT}_{ n,\bar n}
&&=\sum_{ \vec{\bar n}_{ (\;)_1}  \vec n_V } \delta_{\vec{\bar n}_{ (\;)_1}  \vec n_V }
\prod_{i=0}^{D-1}  \delta_{   n^i_{ [s^i_{\cT_1}]^{-1}[ (\;)_1]} \bar n^i_{(\;)_1} }
\prod_{i=0}^{D-1} \delta_{ n^i_V \bar n^i_{s^i_{\cT} [V] } }
\; \delta^{\cT_1\setminus (\;)_1 }_{n,\bar n}\delta^{\cT\setminus V}_{n,\bar n} \crcr
&&= \prod_{i=0}^{D-1} \delta_{n^i_{ [s^i_{\cT_1}]^{-1}[ (\;)_1]}  \bar n^i_{s^i_{\cT} [V] } }
\delta^{\cT_1\setminus (\;)_1 }_{n,\bar n}\delta^{\cT\setminus V}_{n,\bar n}
=\delta^{\cT \star_V \cT_1}_{n,\bar n} \; ,
\eea
hence the SDEs write, for every rooted tree $\cT_1$, with $|R_1|$ non empty branches starting from the root
\bea
&&N^{D-|R_1|}\Big{ \langle } \prod_{ \stackrel{i=0}{\cT_1^i \neq \emptyset}}^{D-1}\text{Tr}_{\cT_1^i}[\bar T, T] \Big{ \rangle }
-N^{D-1} \sum_{\cT}  t_{\cT}  \sum_{V \in \cT } 
\Big{ \langle } \text{Tr}_{\cT \star_{V} \cT_1}[\bar T, T] \Big{ \rangle } = \Big{\langle} \dots \Big{\rangle}
\crcr
&&(-)^{|R_1|}N^{D- D |R_1|} \;  
\Big{(}  \frac{\partial^{|R_1|}} {
\prod_{i, \cT_1^i \neq \emptyset }
\partial t_{\cT_1^i}}  \Big{)} Z
+ \sum_{\cT} t_{\cT} \sum_{V \in \cT }  \frac{\partial}{\partial t_{\cT \star_V \cT_1}} Z = 
\Big{\langle} \dots
 \Big{\rangle} \; ,
\eea
where $\Big{\langle} \dots \Big{\rangle} $ denotes the non melonic terms of eq. \eqref{eq:nonmelonbound}.
Taking into account the definition of $\cL_{\cT_1}$ in eq. \eqref{eq:defcl}, we obtain
\bea
\cL_{\cT_1}Z = \Big{\langle} \dots \Big{\rangle} \Rightarrow \boxed{\lim_{N\to \infty} 
\Big{(}N^{-D} \; \frac{1}{Z} \;\cL_{\cT_1} Z \Big{)}= 0 \;,  \quad \forall \cT_1}\; .
\eea
Recall that $Z = e^{-N^DF(t_{\cB})}$ depends only on class variables $t_{[\cT]}$. At leading 
order in $1/N$ only melonic graph contribute to the free energy $F(t_{\cB})$, 
hence $\lim_{N\to\infty} F(t_{\cB})= F_{\infty} ( [t_{[\cT]}])$.
The SDEs at leading order imply
\bea\label{eq:free}
\boxed {
\prod_{ \stackrel{i=0}{\cT_1^i \neq \emptyset}}^{D-1} 
 \Bigl( \frac{\partial F_{\infty}(t_{[\cT]})}{\partial t_{\cT_1^i}} \Bigr)  - \sum_{\cT} t_{\cT} \sum_{V \in \cT }  
   \frac{\partial F_{\infty}(t_{[\cT]})}{\partial t_{\cT \star_V \cT_1}} =0 \;, \quad \forall \cT_1  }  \; .
\eea 

The most useful way to employ the SDEs is the following. Consider a class function $\tilde Z =e^{-N^{D} \tilde F}$ 
satisfying the constraints at all orders in $N$, $\cL_{\cT_1} \tilde Z =0$. Its free energy in the large $N$ limit,
$\tilde F_{\infty}(t_{[\cT]}) = \lim_{N\to \infty} \tilde F(t_{[\cT]})$ respects eq. \eqref{eq:free}, 
hence $ \tilde F_{\infty}(t_{[\cT]}) = F_{\infty}(t_{[\cT]})$,
that is the $N\to \infty$ limit of $\tilde Z$ and $Z$ coincide.

Note that an SDE at all orders can be derived for the trivial insertion
\bea
 \ \sum_{\vec p} \int  \frac{\delta}{\delta  T_{\vec p } }  
\Big{[} T_{  \vec p }  \;
e^{-N^{D-1} S } \Big{]} =0 \rightarrow 
\Big{(}N^D + \sum_{\cB} |\cB| \;  t_{|\cB| } \;  \frac{\partial}{\partial t_{\cB}}  \Big{)}Z=0\; ,
\eea 
where $|\cB|$ denotes the number of vertices of the bubble $\cB$. The above operator, which at leading order
is $\cL_{(\;)}$, should be identified with the generator of dilations \cite{Dijkgraaf:1990rs}.

\section{Conclusions}\label{sec:conclu}

We have generalized the colored tensor models to colored tensor models with generic interactions, 
derived the Schwinger Dyson equations at leading order and established that (at leading order) the
partition function satisfies a set of constraints forming a Lie algebra.
Much remains to be done in order to fully characterize the critical behavior of the 
colored tensor models. We present below a non exhaustive list of topics one needs to address.

First, although the algebra of melonic bubbles observable closes at leading order in $1/N$, it does not 
closes at all orders (in contrast with matrix models, for which the algebra of loop observables closes 
at all orders). Neither does the algebra of tensor networks corresponding to all bubbles. Indeed, 
if one attempts to derive the full SDEs, one generates terms associated to the addition of lines of color 
$D$ on the $D$-bubbles with colors $0,\dots D-1$. In order to obtain a full set of observables, 
one must also include tensor network operators for the corresponding graphs. The full SDEs 
can be derived, but their algebra is somewhat more involved than the one at leading order. 

A second line of inquiry is to study the algebra of constraints $\cL_{\cT_1}$. 
As colored rooted $D$-ary trees can be indexed in many alternative ways,
in is yet unclear whether this algebra is an entirely new one or some relabeling 
of an already known algebra. One should study in the future its (unitary) 
representations, central extension, etc.. Although we do not yet know what is the
continuum symmetry this algebra encodes, as the generator of dilations is one
of its generators, we expects the continuum theory to be scale invariant.
 
Third, the equation \eqref{eq:free} completely defines the free energy at leading order. 
One can easily write a solution of this equation as a perturbation series in the coupling 
constants. The perturbative solution is ill adapted to the study of $F_{\infty}(t_{[\cT]})$. 
The differential equation \eqref{eq:free} constitutes a much better
starting point for the study of the leading order multi critical behavior of generic colored 
tensor models in arbitrary dimensions.

\section*{Acknowledgements}

The author would like to thank Valentin Bonzom for numerous discussions on matrix models 
and the Virasoro algebra.

Research at Perimeter Institute is supported by the Government of Canada through Industry
Canada and by the Province of Ontario through the Ministry of Research and Innovation.


\begin{thebibliography}{99}


\bibitem{mm}
 F.~David,
  ``A Model Of Random Surfaces With Nontrivial Critical Behavior,''
  Nucl.\ Phys.\  B {\bf 257}, 543 (1985).

\bibitem{david-revueDT}
  F.~David,
  ``Simplicial quantum gravity and random lattices,''
  Les Houches Sum. Sch. 1992:0679-750.
  arXiv:hep-th/9303127.

\bibitem{Di Francesco:1993nw}
  P.~Di Francesco, P.~H.~Ginsparg and J.~Zinn-Justin,
  ``2-D Gravity and random matrices,''
  Phys.\ Rept.\  {\bf 254}, 1 (1995)
  [arXiv:hep-th/9306153].

\bibitem{ambj3dqg}
  J.~Ambjorn, B.~Durhuus and T.~Jonsson,
  ``Three-Dimensional Simplicial Quantum Gravity And Generalized Matrix
  Models,''
  Mod.\ Phys.\ Lett.\  A {\bf 6}, 1133 (1991).

\bibitem{mmgravity}
 M.~Gross,
  ``Tensor models and simplicial quantum gravity in $>$ 2-D,''
  Nucl.\ Phys.\ Proc.\ Suppl.\  {\bf 25A}, 144 (1992).

\bibitem{sasa1}
  N.~Sasakura,
  ``Tensor model for gravity and orientability of manifold,''
  Mod.\ Phys.\ Lett.\  A {\bf 6}, 2613 (1991).

\bibitem{ambjorn-houches94}
  J.~Ambjorn,
  ``Quantization of geometry,''
  Fluctuating Geometries in Statistical Mechanics and Field Theory: Proceedings. Edited by F. David, P. Ginsparg and J. Zinn-Justin. North-Holland, 1996. pp. 77-195. (ISBN 0-444-82294-1).
  arXiv:hep-th/9411179.

\bibitem{ambjorn-book}
  J.~Ambjorn, B.~Durhuus and T.~Jonsson,
  ``Quantum geometry. A statistical field theory approach,''
{\it  Cambridge, UK: Univ. Pr., 1997. (Cambridge Monographs in Mathematical Physics). 363 p}


\bibitem{Boul}
 D.~V.~Boulatov,
  ``A Model of three-dimensional lattice gravity,''
  Mod.\ Phys.\ Lett.\  A {\bf 7}, 1629 (1992)
  [arXiv:hep-th/9202074].

\bibitem{Ooguri:1992eb}
  H.~Ooguri,
  ``Topological lattice models in four-dimensions,''
  Mod.\ Phys.\ Lett.\  A {\bf 7}, 2799 (1992)
  [arXiv:hep-th/9205090].

\bibitem{laurentgft}
  L.~Freidel,
  ``Group field theory: An overview,''
  Int.\ J.\ Theor.\ Phys.\  {\bf 44}, 1769 (2005)
  [arXiv:hep-th/0505016].

\bibitem{quantugeom2}
  D.~Oriti,
  ``The group field theory approach to quantum gravity: some recent results,''
  [arXiv:0912.2441 [hep-th]].

\bibitem{dev1}
 A.~Baratin and D.~Oriti,
  ``Group field theory with non-commutative metric variables,''
  Phys.\ Rev.\ Lett.\  {\bf 105}, 221302 (2010)
  [arXiv:1002.4723 [hep-th]].

\bibitem{dev2}
  W.~J.~Fairbairn and E.~R.~Livine,
  ``3d Spinfoam Quantum Gravity: Matter as a Phase of the Group Field Theory,''
  Class.\ Quant.\ Grav.\  {\bf 24}, 5277 (2007)
  [arXiv:gr-qc/0702125].

\bibitem{dev3}
 A.~Di Mare and D.~Oriti,
  ``Emergent matter from 3d generalised group field theories,''
  Class.\ Quant.\ Grav.\  {\bf 27}, 145006 (2010)
  [arXiv:1001.2702 [gr-qc]].

\bibitem{dev4}
 J.~B.~Geloun, R.~Gurau and V.~Rivasseau,
  ``EPRL/FK Group Field Theory,''
  Europhys.\ Lett.\  {\bf 92}, 60008 (2010)
  [arXiv:1008.0354 [hep-th]].

\bibitem{dev5}
 T.~Krajewski, J.~Magnen, V.~Rivasseau, A.~Tanasa and P.~Vitale,
  ``Quantum Corrections in the Group Field Theory Formulation of the EPRL/FK
  Models,''
  Phys.\ Rev.\  D {\bf 82}, 124069 (2010)
  [arXiv:1007.3150 [gr-qc]].

\bibitem{dev6}
 A.~Tanasa,
  ``Generalization of the Bollob\'as-Riordan polynomial for tensor graphs,''
  arXiv:1012.1798 [math.CO].

\bibitem{Ambjorn:1991wq}
  J.~Ambjorn and S.~Varsted,
  ``Three-dimensional simplicial quantum gravity,''
  Nucl.\ Phys.\  B {\bf 373}, 557 (1992).

\bibitem{Ambjorn:2000dja}
  J.~Ambjorn, J.~Jurkiewicz and R.~Loll,
  ``Nonperturbative 3-D Lorentzian quantum gravity,''
  Phys.\ Rev.\  D {\bf 64}, 044011 (2001)
  [arXiv:hep-th/0011276].

\bibitem{Ambjorn:2005qt}
  J.~Ambjorn, J.~Jurkiewicz and R.~Loll,
  ``Reconstructing the universe,''
  Phys.\ Rev.\  D {\bf 72}, 064014 (2005)
  [arXiv:hep-th/0505154].

\bibitem{Gur3}
   R.~Gurau,
  ``The 1/N expansion of colored tensor models,''
  Annales Henri Poincare {\bf 12}, 829 (2011)
  [arXiv:1011.2726 [gr-qc]].

\bibitem{GurRiv}
   R.~Gurau and V.~Rivasseau,
  ``The 1/N expansion of colored tensor models in arbitrary dimension,''
  arXiv:1101.4182 [gr-qc].

\bibitem{Gur4}
  R.~Gurau,
  ``The complete 1/N expansion of colored tensor models in arbitrary
  dimension,''
  arXiv:1102.5759 [gr-qc].

\bibitem{color}
 R.~Gurau,
  ``Colored Group Field Theory,''
  Commun.\ Math.\ Phys.\  {\bf 304}, 69 (2011)
  [arXiv:0907.2582 [hep-th]].

\bibitem{lost}
  R.~Gurau,
  ``Lost in Translation: Topological Singularities in Group Field Theory,''
  Class.\ Quant.\ Grav.\  {\bf 27}, 235023 (2010)
  [arXiv:1006.0714 [hep-th]].

\bibitem{PolyColor}
  R.~Gurau,
  ``Topological Graph Polynomials in Colored Group Field Theory,''
  Annales Henri Poincare {\bf 11}, 565 (2010)
  [arXiv:0911.1945 [hep-th]].

\bibitem{FreiGurOriti}
  L.~Freidel, R.~Gurau and D.~Oriti,
  ``Group field theory renormalization - the 3d case: power counting of
  divergences,''
  Phys.\ Rev.\  D {\bf 80}, 044007 (2009)
  [arXiv:0905.3772 [hep-th]].

\bibitem{sefu1}
  J.~Magnen, K.~Noui, V.~Rivasseau and M.~Smerlak,
  ``Scaling behavior of three-dimensional group field theory,''
  Class.\ Quant.\ Grav.\  {\bf 26}, 185012 (2009)
  [arXiv:0906.5477 [hep-th]].

\bibitem{sefu2}
  J.~B.~Geloun, J.~Magnen and V.~Rivasseau,
  ``Bosonic Colored Group Field Theory,''
  Eur.\ Phys.\ J.\  C {\bf 70}, 1119 (2010)
  [arXiv:0911.1719 [hep-th]].

\bibitem{sefu3}
   J.~B.~Geloun, T.~Krajewski, J.~Magnen and V.~Rivasseau,
  ``Linearized Group Field Theory and Power Counting Theorems,''
  Class.\ Quant.\ Grav.\  {\bf 27}, 155012 (2010)
  [arXiv:1002.3592 [hep-th]].

 \bibitem{BS1}
  V.~Bonzom and M.~Smerlak,
  ``Bubble divergences from cellular cohomology,''
  Lett.\ Math.\ Phys.\  {\bf 93}, 295 (2010)
  [arXiv:1004.5196 [gr-qc]].

\bibitem{BS2}
  V.~Bonzom and M.~Smerlak,
 ``Bubble divergences from twisted cohomology,''
  arXiv:1008.1476 [math-ph].

\bibitem{Geloun:2011cy}
  J.~B.~Geloun and V.~Bonzom,
  ``Radiative corrections in the Boulatov-Ooguri tensor model: The 2-point
  function,''
  arXiv:1101.4294 [hep-th].

\bibitem{BS3}
  V.~Bonzom and M.~Smerlak,
  ``Bubble divergences: sorting out topology from cell structure,''
  arXiv:1103.3961 [gr-qc].

\bibitem{Carrozza:2011jn}
  S.~Carrozza and D.~Oriti,
  ``Bounding bubbles: the vertex representation of 3d Group Field Theory and
  the suppression of pseudo-manifolds,''
  arXiv:1104.5158 [hep-th].

\bibitem{Brezin:1977sv}
  E.~Brezin, C.~Itzykson, G.~Parisi and J.~B.~Zuber,
  ``Planar Diagrams,''
  Commun.\ Math.\ Phys.\  {\bf 59}, 35 (1978).

\bibitem{'tHooft:1973jz}
  G.~'t Hooft,
  ``A planar diagram theory for strong interactions,''
  Nucl.\ Phys.\  B {\bf 72}, 461 (1974).

\bibitem{Girelli:2010ct}
  F.~Girelli and E.~R.~Livine,
  ``A Deformed Poincare Invariance for Group Field Theories,''
  Class.\ Quant.\ Grav.\  {\bf 27}, 245018 (2010)
  [arXiv:1001.2919 [gr-qc]].

\bibitem{Baratin:2011tg}
  A.~Baratin, F.~Girelli and D.~Oriti,
  ``Diffeomorphisms in group field theories,''
  arXiv:1101.0590 [hep-th].

\bibitem{Ryan:2011qm}
  J.~P.~Ryan,
  ``Tensor models and embedded Riemann surfaces,''
  arXiv:1104.5471 [gr-qc].

\bibitem{Sasakura:2011ma}
  N.~Sasakura,
  ``Tensor models and 3-ary algebras,''
  arXiv:1104.1463 [hep-th].

\bibitem{Sasakura:2011nj}
  N.~Sasakura,
  ``Tensor models and hierarchy of n-ary algebras,''
  arXiv:1104.5312 [hep-th].

\bibitem{Ambjorn:1990ji}
  J.~Ambjorn, J.~Jurkiewicz and Yu.~M.~Makeenko,
  ``Multiloop correlators for two-dimensional quantum gravity,''
  Phys.\ Lett.\  B {\bf 251}, 517 (1990).

\bibitem{Fukuma:1990jw}
  M.~Fukuma, H.~Kawai and R.~Nakayama,
  ``Continuum Schwinger-Dyson Equations and universal structures in 
  two-dimensional quantum gravity,''
  Int.\ J.\ Mod.\ Phys.\  A {\bf 6}, 1385 (1991).

\bibitem{Makeenko:1991ry}
  Yu.~Makeenko,
  ``Loop equations and Virasoro constraints in matrix models,''
  arXiv:hep-th/9112058.

\bibitem{Kazakov:1989bc}
  V.~A.~Kazakov,
  ``The Appearance of Matter Fields from Quantum Fluctuations of 2D Gravity,''
  Mod.\ Phys.\ Lett.\  A {\bf 4}, 2125 (1989).

\bibitem{Gross:1990ay}
  D.~J.~Gross and N.~Miljkovic,
  ``A nonperturbative solution of D = 1 string theory,''
  Phys.\ Lett.\  B {\bf 238}, 217 (1990).

\bibitem{Gross:1990ub}
  D.~J.~Gross and I.~R.~Klebanov,
  ``One-dimensional string theory on a circle,''
  Nucl.\ Phys.\  B {\bf 344}, 475 (1990).

\bibitem{Kazakov:1985ea}
  V.~A.~Kazakov, A.~A.~Migdal and I.~K.~Kostov,
  ``Critical Properties Of Randomly Triangulated Planar Random Surfaces,''
  Phys.\ Lett.\  B {\bf 157}, 295 (1985).

\bibitem{Boulatov:1986jd}
  D.~V.~Boulatov, V.~A.~Kazakov, I.~K.~Kostov and A.~A.~Migdal,
  ``Analytical and Numerical Study of the Model of Dynamically Triangulated
  Random Surfaces,''
  Nucl.\ Phys.\  B {\bf 275}, 641 (1986).

\bibitem{Kazakov:2000pm}
  V.~Kazakov, I.~K.~Kostov and D.~Kutasov,
  ``A matrix model for the two-dimensional black hole,''
  Nucl.\ Phys.\  B {\bf 622}, 141 (2002)
  [arXiv:hep-th/0101011].

\bibitem{Bonzom:2011zz}
  V.~Bonzom, R.~Gurau, A.~Riello and V.~Rivasseau,
  ``Critical behavior of colored tensor models in the large N limit,''
  arXiv:1105.3122 [hep-th].

\bibitem{difrancesco-rect}
  P.~Di Francesco,
  ``Rectangular matrix models and combinatorics of colored graphs,''
  Nucl.\ Phys.\  B {\bf 648}, 461 (2003)
  [arXiv:cond-mat/0208037].

\bibitem{difrancesco-coloringRT}
  P.~Di Francesco, B.~Eynard and E.~Guitter,
  ``Coloring random triangulations,''
  Nucl.\ Phys.\  B {\bf 516}, 543 (1998)
  [arXiv:cond-mat/9711050].

\bibitem{difrancesco-countingRT}
  P.~Di Francesco, B.~Eynard and E.~Guitter,
  ``Counting colored random triangulations,''
  Nucl.\ Phys.\  B {\bf 614}, 519 (2002)
  [arXiv:cond-mat/0206452].

\bibitem{manes-kary-trees}
  K. Manes, A. Sapounakis, I. Tasoulas and P. Tsikouras,
  ``Recursive Generation of k-ary Trees,''
  Journal of Integer Sequences {\bf 12}, 3 (2009).

\bibitem{bianca-finitegp}
  B.~Bahr, B.~Dittrich and J.~P.~Ryan,
  ``Spin foam models with finite groups,''
  arXiv:1103.6264 [gr-qc].

\bibitem{Gal}
  G.~Gallavotti and F.~Nicolo,
  ``Renormalization theory in four-dimensional scalar fields. I,''
  Commun.\ Math.\ Phys.\  {\bf 100}, 545 (1985).

\bibitem{Dijkgraaf:1990rs}
  R.~Dijkgraaf, H.~L.~Verlinde and E.~P.~Verlinde,
  ``Loop equations and Virasoro constraints in nonperturbative 2-D quantum
  gravity,''
  Nucl.\ Phys.\  B {\bf 348}, 435 (1991).


\end{thebibliography}
\end{document}